\documentclass[10pt,english, aps,superscriptaddress,showpacs,floatfix,prl,lengthcheck,floatfix,nofootinbib,twocolumn]{revtex4-2}

\usepackage[T1]{fontenc}
\usepackage[utf8]{inputenc}
\usepackage{verbatim}
\usepackage{bm}
\usepackage{amsmath}
\usepackage{amssymb}
\usepackage{graphicx}
\PassOptionsToPackage{normalem}{ulem}
\usepackage{ulem}

\makeatletter

\newcommand{\lyxmathsym}[1]{\ifmmode\begingroup\def\b@ld{bold}
  \text{\ifx\math@version\b@ld\bfseries\fi#1}\endgroup\else#1\fi}

\usepackage[T1]{fontenc}
\usepackage[utf8]{inputenc} 

\usepackage{braket}

\usepackage{amsmath}

\usepackage{color}
\usepackage{babel}
\usepackage{verbatim}
\usepackage{amsmath}
\usepackage{amssymb}
\usepackage{graphicx}
\usepackage[pdfusetitle,colorlinks,linkcolor=blue,citecolor=blue,urlcolor=blue]{hyperref}
\usepackage{microtype}

\makeatother

\usepackage{babel}
\begin{document}
\title{Semiconductor Wannier equations: \uline{a} real-\uline{t}ime, real-space
\uline{a}pproach to the nonlinear optical response in crys\uline{ta}ls
(ATATA)}
\author{Eduardo B. Molinero}
\email{ebmolinero@gmail.com}

\affiliation{\emph{Instituto de Ciencia de Materiales de Madrid (ICMM), Consejo
Superior de Investigaciones Científicas (CSIC), Sor Juana Inés de
la Cruz 3, 28049 Madrid, Spain}}
\author{Bruno Amorim}
\affiliation{\emph{Centro de Física das Universidades do Minho e do Porto (CF-UM-UP),
Laboratory of Physics for Materials and Emergent Technologies (LaPMET),
Departamento de Física e Astronomia, Faculdade de Ciências, Universidade
do Porto, Rua do Campo Alegre, 4169-007 Porto, Portugal}}
\author{Misha Ivanov}
\affiliation{\emph{Max Born Institute, Max-Born-Straße 2A, 12489, Berlin, Germany}}
\affiliation{\emph{Department of Physics, Humboldt University, Newtonstraße 15,
12489 Berlin, Germany}}
\author{Graham G. Brown}
\affiliation{\emph{Max Born Institute, Max-Born-Straße 2A, 12489, Berlin, Germany}}
\author{Giovanni Cistaro}
\affiliation{\emph{Theory and Simulation of Materials (THEOS), École Polytechnique
Fédérale de Lausanne (EPFL), CH-1015, Lausanne, Switzerland}}
\author{João M. Viana Parente Lopes}
\affiliation{\emph{Centro de Física das Universidades do Minho e do Porto (CF-UM-UP),
Laboratory of Physics for Materials and Emergent Technologies (LaPMET),
Departamento de Física e Astronomia, Faculdade de Ciências, Universidade
do Porto, Rua do Campo Alegre, 4169-007 Porto, Portugal}}
\author{Álvaro Jiménez-Galán}
\affiliation{\emph{Instituto de Ciencia de Materiales de Madrid (ICMM), Consejo
Superior de Investigaciones Científicas (CSIC), Sor Juana Inés de
la Cruz 3, 28049 Madrid, Spain}}
\affiliation{\emph{Max Born Institute, Max-Born-Straße 2A, 12489, Berlin, Germany}}
\author{Pablo San-Jose}
\affiliation{\emph{Instituto de Ciencia de Materiales de Madrid (ICMM), Consejo
Superior de Investigaciones Científicas (CSIC), Sor Juana Inés de
la Cruz 3, 28049 Madrid, Spain}}
\author{Rui E. F. Silva}
\email{ruiefdasilva@gmail.com}

\email{rui.silva@csic.es}

\affiliation{\emph{Instituto de Ciencia de Materiales de Madrid (ICMM), Consejo
Superior de Investigaciones Científicas (CSIC), Sor Juana Inés de
la Cruz 3, 28049 Madrid, Spain}}
\affiliation{\emph{Max Born Institute, Max-Born-Straße 2A, 12489, Berlin, Germany}}
\begin{abstract}
We develop the semiconductor Wannier equations (SWEs), a real-time,
real-space formulation of ultrafast light-matter dynamics in crystals,
by deriving the equations of motion for the electronic reduced density
matrix in a localized Wannier basis. Working in real space removes
the structure-gauge ambiguities that hinder reciprocal-space semiconductor
Bloch equations. Electron--electron interactions are included at
the time-dependent Hartree plus static screened-exchange (TD-HSEX)
level. Decoherence is modeled with three complementary channels: pure
dephasing, population relaxation, and distance-dependent real-space
dephasing; providing physically grounded damping for strong-field
phenomena such as high-harmonic generation. Conceptually, the SWEs
bridge real-space semiclassical intuition with many-body solid-state
optics, offering a numerically robust and gauge-clean alternative
to reciprocal-space approaches for nonlinear optical response and
attosecond spectroscopy in solids.
\end{abstract}
\maketitle

\section{Introduction}

In 2011, the discovery of high-harmonic generation (HHG) in solids
\citep{ghimire2011observation} opened a new frontier in attosecond
science, extending concepts originally developed in atomic and molecular
systems \citep{krausz2009attosecond} to the condensed matter domain
\citep{kruchinin2018colloquium,ghimire2019high,heide2024ultrafast,cavaletto2025attoscience}.
This breakthrough demonstrated that high-order harmonics could be
generated directly from crystalline solids subjected to strong mid-infrared
laser fields, unveiling a new regime of ultrafast light-matter interaction
governed by the interplay of band structure \citep{vampa2015all},
Bloch electron dynamics \citep{ghimire2011observation}, and quantum
coherence \citep{Floss_2018,Brown_2024}.

Since then, high-harmonic spectroscopy has emerged as a powerful tool
to probe electronic and structural properties of solids on femtosecond
and attosecond timescales \citep{heide2024ultrafast}. Applications
range from observation of Bloch oscillations \citep{ghimire2011observation},
optical reconstruction of the bandstructure \citep{vampa2015all},
electron-hole dynamics \citep{hohenleutner2015real}, probing of collective
excitations \citep{molinero2024subcycle}, imaging of van-Hove singularities
\citep{uzan2020attosecond}, topological phase transitions \citep{silva2019topological},
and dynamics in strongly correlated materials \citep{silva2018high,murakami2018high},
to name a few.

The understanding of high-harmonic generation (HHG) in atoms and molecules
is deeply rooted in the semiclassical three-step model \citep{corkum1993plasma,Krause_1992}
and its quantum counterpart, the Lewenstein model \citep{Lewenstein_1994}.
These foundational works underpin the field of attosecond physics,
offering essential physical insight and establishing a clear connection
between HHG and classical electron trajectories \citep{doi:https://doi.org/10.1002/9783527677689.ch7,amini2019symphony}.
\emph{Ab-initio} simulations, especially those solving the time-dependent
Schrödinger equation in real space where electron trajectories can
be directly visualized \citep{Lara_Astiaso_2016}, were instrumental
in identifying electron recollision as the primary mechanism behind
high-harmonic emission in the early days of attoscience \citep{Krause_1992}. 

Several theoretical frameworks have been used to model HHG in solids,
including semiclassical approaches \citep{osika2017wannier,parks2020wannier},
time-dependent density functional theory (TDDFT) \citep{tancogne2017ellipticity,Tancogne_Dejean_2018},
and the semiconductor Bloch equations (SBEs) \citep{lindberg1988effective,Kira_2008,Kira_2011,yue2022introduction,gu2022full,silva2019high}.
Among these approaches, the SBEs in the length gauge, where the coupling
to a homogeneous electric field is described via the scalar potential
$\phi(\boldsymbol{r},t)=-\boldsymbol{r}\cdot\boldsymbol{E}\left(t\right)$,
have proven particularly successful in capturing the underlying physics
and providing valuable insight into the HHG process \citep{yue2022introduction,Parks_2023}.

The advantages of the length gauge formulation of the SBEs when compared
to the velocity gauge are numerous \citep{yue2022introduction,Parks_2023}.
First, less bands are required for the convergence of numerical simulations
when compared to velocity gauge \citep{Yakovlev_2017,yue2020structure}.
This is also the case in atomic and molecular targets, i.e., when
only bound-bound transitions are involved, the length gauge is preferable
over the velocity gauge \citep{Kobe_1979,Bertolino_2022}. Second,
the inclusion of dephasing is most transparently done in the length
gauge \citep{yue2022introduction}. Last, the decomposition of currents
into inter and intraband components is only natural when working in
the length gauge \citep{yue2022introduction}.

Despite its advantages, working in the length gauge presents a significant
challenge due to the fact that the position operator, which appears
in the light--matter interaction term, is ill-defined in a crystal
with periodic boundary conditions \citep{ventura2021nonlinear}. The
standard solution is to define the position operator in the thermodynamic
limit, which requires computing derivatives of the Bloch functions
with respect to Bloch's crystal momentum\citep{blount1962formalisms}.
The gauge freedom in defining the Bloch functions up to a random phase,
which we will refer to as \emph{structure gauge}, poses severe difficulties
to the numerical solution of the SBEs \citep{Jiang_2018,yue2020structure}.
Solutions to this technical problem have been proposed in the literature,
such as the twisted parallel transport gauge \citep{yue2020structure},
the gauge invariant SBEs \citep{Parks_2023,Parks_2024}, and the use
of maximally localized Wannier functions to construct a smooth gauge
in the reciprocal space \citep{silva2019high}. 

In solid state systems, the seminal work of Vampa \emph{et al} \citep{Vampa_2014}
introduced a semiclassical framework for HHG, emphasizing the role
of electron-hole pair trajectories in the process. However, most standard
theoretical tools available to describe electron dynamics in solids,
such as the SBE, are usually formulated in reciprocal space. A theoretical
toolbox formulated in real-space is essential to bridge the gap between
the real-space semiclassical picture and current theoretical methods.

In this work, we present an alternative formulation of the SBEs, the
semiconductor Wannier equations (SWEs), where the dynamical object
propagated in time is the one-electron reduced density matrix, represented
in a real-space basis of localized Wannier orbitals. We will show
that with this approach, the problems that stem from the \emph{structure
gauge }freedom are absent. Furthermore, real-space dephasing to model
solid-state HHG makes the use of unphysically short dephasing times
unnecessary \citep{Brown_2024,Brown_2024josa}. Also, the inclusion
of excitonic effects \citep{cistaro2022theoretical} is substantially
easier than in the SBEs formulation. Finally, in the context of solid-state
attoscience, we demonstrate that SWEs are significantly more efficient
than SBEs opening the way to simulate larger and complex systems.

From a conceptual standpoint, this work opens new avenues for exploring
theoretical quantities in real space to advance the interpretation
of solid-state HHG. It also contributes to bridging the gap between
the concepts developed in atomic and molecular attoscience and their
counterparts in the solid-state regime.

\section{Theoretical Formalism\label{sec:Theoretical_formalism}}

In this section, we will derive the equations of motion that form
the backbone of the semiconductor Wannier equations (SWEs). In the
following, we will use atomic units unless otherwise stated.

\subsection{General formulation}

\subsubsection{Hamiltonian}

We start by writing a general electronic Hamiltonian, $\hat{\mathcal{H}}\left(t\right)$,
that is the sum of a single-particle time-dependent Hamiltonian, $\hat{h}\left(t\right)=\hat{h}^{0}+\hat{U}\left(t\right)$,
and a time-independent electron-electron interaction term, $\hat{V}$.
In a second quantization formalism, we have that

\begin{align}
\hat{\mathcal{H}}\left(t\right) & =\underbrace{\sum_{ij}h_{ij}\left(t\right)c^{\dagger}_{i}c_{j}}_{\hat{h}\left(t\right)}+\underbrace{\frac{1}{2}\sum_{ijkl}V^{ij}_{kl}c^{\dagger}_{i}c^{\dagger}_{j}c_{l}c_{k}}_{\hat{V}}\\
h_{ij}\left(t\right) & =\int d\boldsymbol{r}\phi^{*}_{i}\left(\boldsymbol{r}\right)\hat{h}\left(t\right)\phi_{j}\left(\boldsymbol{r}\right)\\
V^{ij}_{kl} & =\int d\boldsymbol{r}d\boldsymbol{r'}\phi^{*}_{i}\left(\boldsymbol{r}\right)\phi^{*}_{j}\left(\boldsymbol{r'}\right)v_{C}\left(\boldsymbol{r},\boldsymbol{r'}\right)\phi_{l}\left(\boldsymbol{r'}\right)\phi_{k}\left(\boldsymbol{r}\right)
\end{align}
where $c^{\dagger}_{i}$ ($c_{i}$) is the creation (annihilation)
operator for an electron in state $i$ with wavefunction $\phi_{i}\left(\boldsymbol{r}\right)$
and $v_{C}\left(\boldsymbol{r},\boldsymbol{r'}\right)$ is the electron-electron
interaction potential. Since the interaction potential is of the form,
$v_{C}\left(\boldsymbol{r},\boldsymbol{r'}\right)=v_{C}\left(\boldsymbol{r}-\boldsymbol{r'}\right)=v_{C}\left(\boldsymbol{r'}-\boldsymbol{r}\right)$,
the tensor $V^{ij}_{kl}$ has the following symmetries: $V^{ij}_{kl}=V^{ji}_{lk}=\left(V^{kl}_{ij}\right)^{*}$.

In this work, the time-dependent perturbation, $\hat{U}\left(t\right)$,
will be a light-matter interaction term. Under the dipole approximation
and using the electromagnetic length gauge, $\hat{U}\left(t\right)$
can be written as 
\begin{equation}
\hat{U}\left(t\right)=-e\hat{\boldsymbol{r}}\cdot\boldsymbol{E}\left(t\right),\label{eq:dipole_interaction}
\end{equation}
where $e=-1$ is the electron charge, $\boldsymbol{E}\left(t\right)$
is the applied homogeneous electric field and $\hat{\boldsymbol{r}}$
is the position operator. 

\subsubsection{Observables and the one-electron reduced density matrix}

In most of the cases, we are interested in calculating observables
that are single-particle operators, which have the general form $\hat{O}=\sum_{ij}O_{ij}c^{\dagger}_{i}(t)c_{j}(t)$
(from hereinafter we will be using the Heisenberg picture). The expectation
value of a single-particle operator can be written as $\left\langle \hat{O}(t)\right\rangle =\sum_{ij}O_{ij}\rho_{ji}(t)=\mathrm{tr}\left(O\rho(t)\right)$,
where
\begin{equation}
\rho_{ij}(t)\equiv\left\langle c^{\dagger}_{j}(t)c_{i}(t)\right\rangle 
\end{equation}
is the one-electron reduced density matrix (1RDM). Note that the indexes
are switched.

Regarding optical response, two of the most crucial and important
observables are the position and velocity operators. The position
operator is just expressed as
\begin{align}
\hat{\boldsymbol{r}} & =\sum_{ij}\boldsymbol{r}_{ij}c^{\dagger}_{i}c_{j}\\
\boldsymbol{r}_{ij} & =\int d\boldsymbol{r'}\phi^{*}_{i}\left(\boldsymbol{r'}\right)\boldsymbol{r'}\phi_{j}\left(\boldsymbol{r'}\right).
\end{align}
The velocity operator, $\hat{\boldsymbol{v}}$, can be calculated
using the Heisenberg equations of motion 
\begin{equation}
\hat{\boldsymbol{v}}=-i\left[\hat{\boldsymbol{r}},\hat{\mathcal{H}}\left(t\right)\right].
\end{equation}
The electron-electron interaction term, $\hat{V}$, is an operator
that only depend on the position of the electrons, therefore, it commutes
with $\hat{\boldsymbol{r}}$. Equivalently, this reflects the fact
that internal forces do not modify the total current, in agreement
with Newton’s third law. Furthermore, if the time-dependent perturbation,
$\hat{U}\left(t\right)$, also commutes with the position, as it is
the case when working under the dipole approximation in the length
gauge, the velocity operator can be written as 
\begin{align}
\hat{\boldsymbol{v}} & =\sum_{ij}\boldsymbol{v}_{ij}c^{\dagger}_{i}c_{j}=-i\left[\hat{\boldsymbol{r}},\hat{h}^{0}\right]\label{eq:vel_operator}
\end{align}
where $\boldsymbol{v}_{ij}=-i\sum_{k}\left(\boldsymbol{r}_{ik}h^{0}_{kj}-h^{0}_{ik}\boldsymbol{r}_{kj}\right)$
are the velocity matrix elements and $\left\langle \hat{\boldsymbol{v}}\right\rangle =\mathrm{tr}\left(\boldsymbol{v}\rho\right)$.
It is important to emphasize that the calculation of the velocity
operator is done without assuming any approximations. In the following,
we will approximate the equations of motion by performing a mean-field
decoupling, effectively adding a self-energy term, $\Sigma$, to $h^{0}$.
One might be tempted to include in the definition of the velocity
operator this self-energy term, $\Sigma$. This would be incorrect,
however Eq. (\ref{eq:vel_operator}) was derived without any approximation,
and it is the correct expression for $\hat{\boldsymbol{v}}$ even
in the presence of $\Sigma$. Recomputing the velocity as $-i\left[\boldsymbol{\hat{r}},\hat{h}^{0}+\Sigma\right]$
would therefore amount to double counting the interaction effect:
the interaction has already been accounted for once in the exact derivation
of the current operator, and $\Sigma$ is only introduced afterwards
as an approximate closure for the dynamics.

\subsubsection{Equations of motion for the one-electron reduced density matrix}

Using the Heisenberg equation, we will determine the equation of motion
for the 1RDM, $\rho_{ij}(t)=\left\langle c^{\dagger}_{j}(t)c_{i}(t)\right\rangle $.
We have used \textbf{QuantumAlgebra.jl} \citep{Sanchez-Barquilla2020,QuantumAlgebra.jl}
to deal with the quantum operator algebra. We have that (omitting
the time-argument for sinplicity)
\begin{align}
-i\frac{d\left(c^{\dagger}_{j}c_{i}\right)}{dt} & =\left[\hat{\mathcal{H}}\left(t\right),c^{\dagger}_{j}c_{i}\right]\\
-i\frac{d\rho_{ij}}{dt} & =\left\langle \left[\hat{\mathcal{H}}\left(t\right),c^{\dagger}_{j}c_{i}\right]\right\rangle \label{eq:1RDM}\\
 & =\sum_{\alpha}h_{\alpha j}\left(t\right)\rho_{i\alpha}-\sum_{\alpha}h_{i\alpha}\left(t\right)\rho_{\alpha j}\nonumber \\
 & -\frac{1}{2}\sum_{\alpha\beta\gamma}V^{\alpha\beta}_{\gamma j}\left\langle c^{\dagger}_{\alpha}c^{\dagger}_{\beta}c_{\gamma}c_{i}\right\rangle \nonumber \\
 & +\frac{1}{2}\sum_{\alpha\beta\gamma}V^{\alpha\beta}_{j\gamma}\left\langle c^{\dagger}_{\alpha}c^{\dagger}_{\beta}c_{\gamma}c_{i}\right\rangle \nonumber \\
 & +\frac{1}{2}\sum_{\alpha\beta\gamma}V^{\alpha i}_{\beta\gamma}\left\langle c^{\dagger}_{\alpha}c^{\dagger}_{j}c_{\beta}c_{\gamma}\right\rangle \nonumber \\
 & -\frac{1}{2}\sum_{\alpha\beta\gamma}V^{i\alpha}_{\beta\gamma}\left\langle c^{\dagger}_{\alpha}c^{\dagger}_{j}c_{\beta}c_{\gamma}\right\rangle .\nonumber 
\end{align}
We arrive to an equation of motion for the 1RDM that involves the
computation and evolution of the expectation value of four-operators,
$\left\langle c^{\dagger}_{i}c^{\dagger}_{j}c_{l}c_{k}\right\rangle $.
If we proceed to compute the equation of motion for the mean value
of four-operators, we will get equations involving mean values of
six-operators. If we proceed further, we end up with an infinite hierarchy
of operators, also know as\textbf{ }Bogoliubov--Born--Green--Kirkwood--Yvon
(BBGKY) hierarchy \citep{Kira_2008,Kira_2011,Sanchez-Barquilla2020}.
In the following, we will use a mean-field decoupling to obtain a
close set of equations for the 1RDM.

\subsubsection{Mean-field decoupling}

To obtain a closed set of equations, we can apply a mean-field decoupling
to the $\left\langle c^{\dagger}_{i}c^{\dagger}_{j}c_{l}c_{k}\right\rangle $
terms, equivalent to a Hartree-Fock approximation. Specifically, we
will approximate \citep{bruus2004many,ribeiro2023excitonic}
\begin{align}
\left\langle c^{\dagger}_{i}c^{\dagger}_{j}c_{l}c_{k}\right\rangle  & \approx\left\langle c^{\dagger}_{i}c_{k}\right\rangle \left\langle c^{\dagger}_{j}c_{l}\right\rangle -\left\langle c^{\dagger}_{i}c_{l}\right\rangle \left\langle c^{\dagger}_{j}c_{k}\right\rangle \nonumber \\
 & \approx\rho_{ki}\rho_{lj}-\rho_{li}\rho_{kj}.
\end{align}
Note that in this decoupling, we neglect anomalous terms, $\left\langle c^{\dagger}_{i}c^{\dagger}_{j}\right\rangle $
and $\left\langle c_{i}c_{j}\right\rangle $, that may be relevant
to superconductivity \citep{lobo2024exponential}. Replacing this
decoupling in the equation of motion for the 1RDM, Eq. (\ref{eq:1RDM}),
we obtain the equation of motion for the 1RDM within the time-dependent
Hartree-Fock approximation

\begin{equation}
i\frac{d\rho_{ij}}{dt}=\left[h\left(t\right),\rho\right]_{ij}+\left[\Sigma^{F}\left[\rho\right],\rho\right]_{ij}+\left[\Sigma^{H}\left[\rho\right],\rho\right]_{ij}\label{eq:eom_hf}
\end{equation}
where $\Sigma^{F}$ and $\Sigma^{H}$ are the Fock and Hartree self-energies
(mean-field potentials) and are defined as
\begin{align}
\Sigma^{F}\left[\rho\right]_{ij} & =-\sum_{kl}V^{il}_{kj}\rho_{kl}\label{eq:Fock_self}\\
\Sigma^{H}\left[\rho\right]_{ij} & =\sum_{kl}V^{il}_{jk}\rho_{kl}.
\end{align}

Up to this point, $\hat{h}^{0}$ has been treated as a purely non-interacting
Hamiltonian. However, in practical calculations, $\hat{h}^{0}$, and
therefore the equilibrium RDM\footnote{The equilibrium RDM is just $\rho^{0}=F_{\mu,T}\left(h^{0}\right)$,
being $F_{\mu,T}$ the Fermi function for chemical potential $\mu$
and temperature $T$. In the eigenbasis of $\hat{h}^{0}\left|\varepsilon_{i}\right\rangle =\varepsilon_{i}\left|\varepsilon_{i}\right\rangle $,
$\rho^{0}_{ij}=\delta_{ij}F_{\mu,T}\left(\varepsilon_{i}\right)$.}, $\rho^{0}$, are obtained from either tight-binding parametrizations
of the experimental equilibrium dispersion or from different levels
of \emph{ab-initio} calculations, such as $GW$ \citep{martinbookinteracting}
or hybrid DFT calculations \citep{martin2004electronic}. In both
cases, the resulting $\hat{h}^{0}$ already includes some level of
electron-electron interactions, in the form of a single-particle self-energy
$\Sigma_{0}$ at equilibrium. Therefore, in the equation of motion
Eq. (\ref{eq:eom_hf}) we must subtract the equilibrium ground-state
self-energy, $\Sigma_{0}=\Sigma^{F}\left[\rho_{0}\right]+\Sigma^{H}\left[\rho_{0}\right]$,
to avoid double counting.

In the present formulation, $\hat{h}^{0}$ is used as an effective
single-particle Hamiltonian for the active subspace. Thus, quasiparticle-level
renormalizations of the band energies, including band-gap renormalization,
are assumed to be incorporated, at least effectively, in the input
band structure. The interaction terms treated explicitly below should
therefore be understood as residual screened interactions within the
reduced subspace. 

Secondly, the Fock self-energy neglects important screening effects,
that are usually taken into account within a random-phase approximation
(RPA) \citep{bruus2004many,martinbookinteracting}. As a result, the
interaction potential $V$ in Eq. (\ref{eq:Fock_self}) is often replaced
with $W$, the RPA-screened interaction potential. Note that, this
argument does not apply to the Hartree self-energy since RPA screening
would result in diagrams that are not single-particle irreducible,
leading to double-counting of screening. The screened interaction
$W$ is, in general, a dynamical quantity, $W(\omega)$, as it depends
on the state of the system. However, in order to keep the equations
of motion local in time we will employ a common approximation, known
as static exchange (SEX), by taking the static limit, i.e. $W\approx W(\omega\rightarrow0)$
\citep{martinbookinteracting}. Therefore, our final equation of motion
will read 
\begin{align}
i\frac{d\rho_{ij}}{dt} & =\left[h\left(t\right)+\Sigma^{HSEX}\left[\rho\right]-\Sigma_{0},\rho\right]_{ij},\label{eq:TD-HSEX}\\
\Sigma^{SEX}\left[\rho\right]_{ij} & =-\sum_{kl}W^{il}_{kj}\rho_{kl},\\
\Sigma^{HSEX}\left[\rho\right]_{ij} & =\Sigma^{SEX}\left[\rho\right]_{ij}+\Sigma^{H}\left[\rho\right]_{ij},\\
\Sigma_{0} & =\Sigma^{HSEX}\left[\rho^{0}\right].
\end{align}
The $\Sigma_{0}$ term is chosen to be the full self-energy at equilibrium,
regardless of the level of approximation used to compute $h^{0}$
and $\rho^{0}$. This will ensure that $\rho^{0}$, the equilibrium
1RDM, will remain a stationary solution of the equations of motion,
in the absence of driving. This approach is sometimes known in the
literature as TD-BSE \citep{Attaccalite_2011}, TD-aGW \citep{tumakov2026real}
or TD-HSEX \citep{sangalli2018ab,sangalli2021excitons,sangalli2023exciton}.

The specific form of $V^{ij}_{kl}$ and $W^{ij}_{kl}$ can be simplified
when working with a localized basis, as occur in a tight binding formulation.
Specifically, assuming that the orbitals $\phi^{*}_{i}\left(\boldsymbol{r}\right)$
are strongly localized around the center of the orbital, $\bm{\tau}_{i}=\left\langle i\left|\boldsymbol{r}\right|i\right\rangle $,
we can approximate
\begin{align}
V^{ij}_{kl} & =\int d\boldsymbol{r}d\boldsymbol{r'}\phi^{*}_{i}\left(\boldsymbol{r}\right)\phi^{*}_{j}\left(\boldsymbol{r'}\right)V\left(\boldsymbol{r}-\boldsymbol{r'}\right)\phi_{l}\left(\boldsymbol{r'}\right)\phi_{k}\left(\boldsymbol{r}\right)\nonumber \\
 & \approx V\left(\boldsymbol{\tau}_{i}-\boldsymbol{\tau}_{j}\right)\int d\boldsymbol{r}d\boldsymbol{r'}\phi^{*}_{i}\left(\boldsymbol{r}\right)\phi^{*}_{j}\left(\boldsymbol{r'}\right)\phi_{l}\left(\boldsymbol{r'}\right)\phi_{k}\left(\boldsymbol{r}\right),
\end{align}
which is a valid approximation, provided $V\left(\boldsymbol{r}-\boldsymbol{r'}\right)$
does not vary strongly within typical size of an orbital. If we further
recall the ortogonality of Wannier orbitals,  $\int d\boldsymbol{r}\phi^{*}_{i}\left(\boldsymbol{r}\right)\phi_{k}\left(\boldsymbol{r}\right)=\delta_{ik}$,
we obtain

\begin{equation}
V^{ij}_{kl}\approx\delta_{ik}\delta_{jl}V\left(\boldsymbol{\tau}_{i}-\boldsymbol{\boldsymbol{\tau}}_{j}\right)\label{eq:ultraloc_1}
\end{equation}
where $\boldsymbol{\boldsymbol{\tau}}_{i},\,\boldsymbol{\boldsymbol{\tau}}_{j}$
are the centers of the orbitals $i$ and $j$. In the same way, 
\begin{equation}
W^{ij}_{kl}\approx\delta_{ik}\delta_{jl}W\left(\boldsymbol{\tau}_{i}-\boldsymbol{\tau}_{j}\right).\label{eq:ultra_loc_2}
\end{equation}
We refer to this as the ultra-localized orbital approximation \citep{ribeiro2023excitonic}.
Within this approximation, we can then calculate $\Sigma^{SEX}\left[\rho\right]$
and $\Sigma^{H}\left[\rho\right]$ as 
\begin{align}
\Sigma^{SEX}\left[\rho\right]_{ij} & \approx-W\left(\boldsymbol{\tau}_{i}-\boldsymbol{\boldsymbol{\tau}}_{j}\right)\rho_{ij},\\
\Sigma^{H}\left[\rho\right]_{ij} & \approx\delta_{ij}\sum_{k}\rho_{kk}V\left(\boldsymbol{\tau}_{i}-\boldsymbol{\boldsymbol{\tau}}_{k}\right).
\end{align}
When working with spin-degenerate models, we must multiply the Hartree
term, $\Sigma^{H}\left[\rho\right]$, by 2 since it depends on the
total charge density. Therefore,
\begin{equation}
\Sigma^{H}\left[\rho\right]_{ij}\approx\delta_{ij}\sum_{k}s_{k}\rho_{kk}V\left(\boldsymbol{\tau}_{i}-\boldsymbol{\boldsymbol{\tau}}_{k}\right)
\end{equation}
where $s_{k}$ takes into account the spin-degeneracy of the $k$
orbital. 

We emphasize that the ultra-localized orbital approximation is not
an essential component of the SWEs formalism, but an additional approximation
adopted here to simplify the evaluation of the two-electron integrals
entering the $\Sigma^{HSEX}$ self-energy. The SWEs can be formulated
using the full interaction matrix elements, $V^{ij}_{kl}$ and $W^{ij}_{kl}$,
although at a higher computational cost. The approximation is expected
to be reasonable when the Wannier orbitals are sufficiently localized
and the interaction potentials do not vary significantly over their
spatial extent.

\subsubsection{Decoherence\label{subsec:Decoherence}}

The equation of motion derived in the previous subsection, Eq. (\ref{eq:TD-HSEX}),
is purely coherent, with no dissipation effects. However, decoherence
plays a major role in the modelization of electronic dynamics in solids
under intense lasers, particularly in high harmonic generation \citep{Vampa_2014,Floss_2018,Brown_2024,Brown_2024josa}.
To that purpose, we must include decoherence in our framework. We
will include three forms of decoherence: pure dephasing, relaxation
and real-space dephasing. 

In the seminal work by Vampa \emph{et al }\citep{Vampa_2014}, it
was found that HHG spectrum obtained by solving the SBEs in ZnO was
very noisy and agreement with experiments was only achieved if ultrashort
dephasing times of few femtoseconds, $T_{2}\equiv\gamma^{-1}_{D}$,
were introduced in the simulations. Pure dephasing is introduced in
the basis that diagonalizes the unperturbed Hamiltonian, $\hat{h}^{0}\left|\varepsilon_{i}\right\rangle =\varepsilon_{i}\left|\varepsilon_{i}\right\rangle $.
The pure dephasing term in the equation of motion, in the eigenbasis
of $\hat{h}^{0}$, reads
\begin{equation}
\mathcal{L}_{D}\left[\rho\right]_{ij}=\gamma_{D}\left(\delta_{ij}-1\right)\rho_{ij}.
\end{equation}

Despite pure dephasing being commonly used in numerical simulations
of HHG in solids, it does not account for population relaxation. To
that purpose, within the relaxation time approximation and assuming
that the relaxation rate is much smaller than the driving frequency,
we include relaxation to thermal equilibrium, $\rho^{0}$ \citep{terada2024limitations,cox2014electrically,sato2021nonlinear}.
The relaxation term reads as 
\begin{equation}
\mathcal{L}_{r}\left[\rho\right]=-\gamma_{r}\left(\rho-\rho_{0}\right),\label{eq:relaxation_term}
\end{equation}
where $\gamma_{r}\equiv\left(2T_{r}\right)^{-1}$ is the relaxation
rate. 

Recently, Brown\emph{ et al.} \citep{Brown_2024,Brown_2024josa} introduced
an additional term that implements real-space dephasing where the
rate of dephasing depends on the distance between orbitals. In this
specific case, we must work in a basis where the single-particle orbitals
are well localized. Having $\boldsymbol{\tau}_{i}$ as the center
of state $i$, we have that the real-space dephasing term reads as
\begin{equation}
\mathcal{L}_{rs}\left[\rho\right]_{ij}=-\gamma_{rs}\left(\left|\boldsymbol{\tau}_{i}-\boldsymbol{\tau}_{j}\right|\right)\cdot\left(\rho-\rho^{0}\right)_{ij},
\end{equation}
where $\gamma_{rs}\left(\left|\boldsymbol{\tau}_{i}-\boldsymbol{\tau}_{j}\right|\right)$
is a dephasing rate that depends on the distance between orbitals
$i$ and $j$ and is zero at the origin, $\gamma_{rs}\left(0\right)=0$.
The functional form that will be used for $\gamma_{rs}$ is a polynomial
function that only acts on coherences between states whose distance
is greater than $R_{rs,cut}$, i.e.
\begin{equation}
\gamma_{rs}\left(r\right)=\begin{cases}
0 & r<R_{rs,cut}\\
\beta_{rs}\left(r-R_{rs,cut}\right)^{\alpha_{rs}} & r\ge R_{rs,cut}
\end{cases}\label{eq:rs_dephasing_form}
\end{equation}
where $\alpha_{rs}$ is the degree of the polynomial. This form of
real-space dephasing is similar to the one used in \citep{Brown_2024}
and resembles the complex absorbing potentials used in atomic and
molecular strong field calculations \citep{scrinzi2010infinite,richter2013role,pedatzur2015attosecond,silva2016even}.

\subsection{Crystalline systems}

We are now in position to lay down the equations of motion for the
electronic RDM in a crystalline solid in real-space. Let us consider
an infinite crystalline solid with the Bravais lattice 
\begin{equation}
\boldsymbol{R}=\sum^{D}_{i=1}n_{i}\boldsymbol{a}_{i},\ n_{i}\in\mathbb{Z},
\end{equation}
where $D$ is the dimension of the system, $\boldsymbol{a}_{i}$ are
the primitive lattice vectors and $\boldsymbol{b}_{i}$ denotes the
primitive reciprocal lattice vectors such that $\boldsymbol{a}_{i}\cdot\boldsymbol{b}_{j}=2\pi\delta_{ij}$.
Let us consider a set of $M$ localized Wannier spin-orbitals in each
cell $w_{\alpha}\left(\boldsymbol{r}-\boldsymbol{R}\right)=\left\langle \boldsymbol{r}|\boldsymbol{R}\alpha\right\rangle $.
These Wannier orbitals are characterized by two indexes $\left(\boldsymbol{R},\alpha\right)$,
with $\mathbf{R}$ the center of the orbital and $\alpha$ the orbital
type, and form an orthonormal basis, i.e., $\left\langle \boldsymbol{R}'\beta|\boldsymbol{R}\alpha\right\rangle =\delta_{\alpha\beta}\delta_{\boldsymbol{R}',\boldsymbol{R}}$.
Throughout this work, we use the terms “Wannier orbitals” and “localized
orbitals” interchangeably. We will denote $\rho_{\alpha\beta}\left(\boldsymbol{R}',\boldsymbol{R}\right)\equiv\left\langle c^{\dagger}_{\boldsymbol{R}\beta}c_{\boldsymbol{R}'\alpha}\right\rangle $
and $\rho_{\alpha\beta}\left(\boldsymbol{R}\right)\equiv\rho_{\alpha\beta}\left(\boldsymbol{0},\boldsymbol{R}\right)$.
As a consequence of the periodicity of our system, the equilibrium
state, $\rho^{0}$, obeys 
\begin{equation}
\rho^{0}_{\alpha\beta}\left(\boldsymbol{R}-\boldsymbol{R}'\right)=\rho^{0}_{\alpha\beta}\left(\boldsymbol{R}',\boldsymbol{R}\right).
\end{equation}
This is a consequence of Bloch's theorem, that in reciprocal space
only allows for non-vanishing coherences between states with the same
$\boldsymbol{k}$. 

Two basic quantities in the problem of light-matter interaction are
the unperturbed Hamiltonian, $\hat{h}^{0}$, and the position operator,
$\hat{\boldsymbol{r}}$. Their matrix elements can be either calculated
from electronic structure calculations followed by a wannierization
procedure \citep{marzari2012maximally} or be defined in the context
of a phenomenological tight-binding model. For the description of
light-matter interaction under the dipole approximation it is necessary
to have the knowledge of their matrix elements
\begin{align}
h^{0}_{\alpha\beta}\left(\boldsymbol{R}\right)\equiv & \left\langle \boldsymbol{0}\alpha\left|\hat{h}^{0}\right|\boldsymbol{R}\beta\right\rangle ,\\
\boldsymbol{r}_{\alpha\beta}\left(\boldsymbol{R}\right)\equiv & \left\langle \boldsymbol{0}\alpha\left|\hat{\boldsymbol{r}}\right|\boldsymbol{R}\beta\right\rangle .
\end{align}
The unperturbed Hamiltonian is invariant under lattice translations,
i.e., $\left\langle \boldsymbol{0}\alpha\left|\hat{h}^{0}\right|\left(\boldsymbol{R}-\boldsymbol{R}'\right)\beta\right\rangle =\left\langle \boldsymbol{R}'\alpha\left|\hat{h}^{0}\right|\boldsymbol{R}\beta\right\rangle $. 

On the other hand, the position operator is not invariant under lattice
translations and transforms as
\begin{equation}
\left\langle \boldsymbol{R}'\alpha\left|\hat{\boldsymbol{r}}\right|\boldsymbol{R}\beta\right\rangle =\boldsymbol{r}_{\alpha\beta}\left(\boldsymbol{R}-\boldsymbol{R}'\right)+\delta_{\boldsymbol{R}',\boldsymbol{R}}\delta_{\alpha\beta}\boldsymbol{R}.\label{eq:r_operator_transform}
\end{equation}
At first glance, the second term in Eq. (\ref{eq:r_operator_transform})
might seem problematic as it breaks the periodicity of our problem.
However, under the dipole approximation, the reduced density matrix
in real-space is an object that remains invariant under lattice translations,
i.e. $\rho_{\alpha\beta}\left(\boldsymbol{R}-\boldsymbol{R}'\right)=\rho_{\alpha\beta}\left(\boldsymbol{R}',\boldsymbol{R}\right)$.
Lets take a closer look at the equation of motion for the RDM by only
taking into account the problematic $\delta_{\boldsymbol{R}',\boldsymbol{R}}\delta_{\alpha\beta}\boldsymbol{R}$
term
\begin{equation}
i\frac{d\rho_{\alpha\beta}\left(\boldsymbol{R}',\boldsymbol{R}\right)}{dt}=-\boldsymbol{E}\left(t\right)\cdot\left(\boldsymbol{R}-\boldsymbol{R}'\right)\rho_{\alpha\beta}\left(\boldsymbol{R}',\boldsymbol{R}\right).
\end{equation}
Taking into account that in our initial state, $\rho^{0}_{\alpha\beta}\left(\boldsymbol{R}',\boldsymbol{R}\right)=\rho^{0}_{\alpha\beta}\left(\boldsymbol{0},\boldsymbol{R}-\boldsymbol{R}'\right)$,
we can see that the previous term in the equation of motion will maintain
the lattice invariance of the 1RDM. In the following, we will always
assume implicitly this property of the 1RDM, i.e. 
\begin{equation}
\rho_{\alpha\beta}\left(\boldsymbol{R}\right)=\rho_{\alpha\beta}\left(\boldsymbol{R}',\boldsymbol{R}+\boldsymbol{R}'\right),
\end{equation}
and it is this quantity that will be our central dynamical object. 

Both self-energy terms, $\Sigma^{H}$ and $\Sigma^{SEX}$, will also
be invariant under lattice translations. In the following, we employ
the ultra-localized orbital approximation introduced in Eqs. (\ref{eq:ultraloc_1})
and (\ref{eq:ultra_loc_2}). Under this approximation, the SEX self-energy
term takes the form
\begin{align}
\Sigma^{SEX}\left[\rho\right]_{\alpha\beta}\left(\boldsymbol{R}\right) & =-W\left(\boldsymbol{\tau}_{\alpha}-\left(\boldsymbol{\tau}_{\beta}+\boldsymbol{R}\right)\right)\rho_{\alpha\beta}\left(\boldsymbol{R}\right),
\end{align}
where $\boldsymbol{\tau}_{\alpha}=\left\langle \boldsymbol{0}\alpha\left|\hat{\boldsymbol{r}}\right|\boldsymbol{0}\alpha\right\rangle $
is the center of the $\alpha$ Wannier function at the $\boldsymbol{0}$
unit cell. The Hartree self-energy can be written as 
\begin{align}
\Sigma^{H}\left[\rho\right]_{\alpha\beta}\left(\boldsymbol{R}\right) & =\delta_{\alpha\beta}\delta_{\boldsymbol{0},\boldsymbol{R}}\sum_{\gamma}\rho_{\gamma\gamma}\left(\boldsymbol{0}\right)X_{\gamma\alpha},\\
X_{\gamma\alpha} & \equiv\sum_{\boldsymbol{R'}}s_{\gamma}V\left(\boldsymbol{\tau}_{\alpha}-\left(\boldsymbol{\tau}_{\gamma}+\boldsymbol{R'}\right)\right)
\end{align}
where $s_{\gamma}$ takes into account the spin-degeneracy of the
orbital $\gamma$.

\subsubsection{Equations of motion in Wannier basis}

We are now in position to derive the equations of motion for $\rho^{L}_{\alpha\beta}\left(\boldsymbol{R}\right)$,
where the upperscript $L$ stands for length gauge, as defined by
Eq. (\ref{eq:dipole_interaction}). In the following, we will omit
the indexes for the Wannier orbitals implying that we are referring
to a $M\times M$ matrix. The resulting equation of motion is

\begin{align}
i\partial_{t}\rho^{L}\left(\boldsymbol{R}\right) & =\sum_{\boldsymbol{R'}}\left[h^{0}\left(\boldsymbol{R}'\right),\rho^{L}\left(\boldsymbol{R}-\boldsymbol{R'}\right)\right]\nonumber \\
 & +\boldsymbol{E}\left(t\right)\cdot\sum_{\boldsymbol{R'}}\left[\boldsymbol{r}\left(\boldsymbol{R}'\right),\rho^{L}\left(\boldsymbol{R}-\boldsymbol{R'}\right)\right]\nonumber \\
 & -\boldsymbol{E}\left(t\right)\cdot\boldsymbol{R}\rho^{L}\left(\boldsymbol{R}\right)\nonumber \\
 & +\sum_{\boldsymbol{R'}}\left[\Sigma^{HSEX}\left[\rho^{L}\right]\left(\boldsymbol{R}'\right),\rho^{L}\left(\boldsymbol{R}-\boldsymbol{R'}\right)\right]\nonumber \\
 & -\sum_{\boldsymbol{R'}}\left[\Sigma_{0}\left(\boldsymbol{R}'\right),\rho^{L}\left(\boldsymbol{R}-\boldsymbol{R'}\right)\right]\nonumber \\
 & +i\mathcal{L}_{incoh}\left[\rho^{L}\right]\left(\boldsymbol{R}\right),\label{eq:eom_length_gauge}
\end{align}
where we group all incoherent terms, discussed in \ref{subsec:Decoherence},
into
\begin{align}
\mathcal{L}_{incoh} & =\mathcal{L}_{r}+\mathcal{L}_{rs}+\mathcal{L}_{D}.
\end{align}
The explicit formulation of the incoherent terms in the case of a
periodic crystal will be shown later. 

\subsubsection{Length gauge and Peierls gauge}

The term $\boldsymbol{E}\left(t\right)\cdot\boldsymbol{R}\rho^{L}\left(\boldsymbol{R}\right)$
in Eq. (\ref{eq:eom_length_gauge}) clearly diverges in a crystal
and this poses severe numerical difficulties. These difficulties might
be solved by working with sufficiently bigger lattices, since it should
be expected that $\rho\left(\boldsymbol{R}\right)\rightarrow0$ when
$\left|\boldsymbol{R}\right|\rightarrow\infty$. At the same time,
dealing with the $\boldsymbol{E}\left(t\right)\cdot\boldsymbol{R}$
term for big $\boldsymbol{R}$ will require a very fine temporal resolution.
However, one might find a more elegant solution to this problem. In
a reciprocal space formulation, the position operator is expressed
as a gradient in $k$-space \citep{blount1962formalisms,ventura2017gauge,ventura2021nonlinear}.
One might numerically discretize the gradient, as done in previous
works \citep{marzari1997maximally,cistaro2022theoretical}. An alternative
way to deal with the gradient of the density matrix in the reciprocal
space is to move to the accelerated basis \citep{yue2022introduction}.
In our real-space approach, a time-dependent translation in reciprocal
space is represented as a time-dependent phase shift in real space.
At the level of the reduced density matrix, this is the real-space
counterpart of the Houston-basis transformation \citep{yue2022introduction}
in reciprocal-space formulations. Therefore, one might define $\rho^{P}\left(\boldsymbol{R}\right)$,
the real-space 1RDM in the Peierls gauge, as 
\begin{equation}
\rho^{P}\left(\boldsymbol{R}\right)=\rho^{L}\left(\boldsymbol{R}\right)\phi_{P}\left(t,-\boldsymbol{R}\right)\label{eq:lenght_peierls_transformation}
\end{equation}
where $\phi_{P}\left(t,\boldsymbol{R}\right)=\exp\left(-i\boldsymbol{A}_{L}\left(t\right)\cdot\boldsymbol{R}\right)$
is the Peierls phase between unit cells and $\boldsymbol{A}_{L}\left(t\right)=-\int^{t}\boldsymbol{E}\left(t'\right)dt'$
is the laser vector potential. We might define $\mathcal{G}^{P\rightarrow L}$
as the transformation from the length gauge to the Peierls gauge as
\begin{equation}
\mathcal{G}^{L\rightarrow P}\left[\rho^{L}\left(\boldsymbol{R}\right)\right]\equiv\rho^{P}\left(\boldsymbol{R}\right)=\rho^{L}\left(\boldsymbol{R}\right)\phi_{P}\left(t,-\boldsymbol{R}\right)
\end{equation}
and the corresponding inverse transformation, $\mathcal{G}^{P\rightarrow L}$.
The equation of motion for $\rho^{P}\left(\boldsymbol{R}\right)$
is \begin{widetext}

\begin{align}
i\partial_{t}\rho^{P}\left(\boldsymbol{R}\right) & =\sum_{\boldsymbol{R'}}\phi_{P}\left(t,-\boldsymbol{R'}\right)\left[h^{0}\left(\boldsymbol{R}'\right)+\boldsymbol{E}\left(t\right).\boldsymbol{r}\left(\boldsymbol{R}'\right),\rho^{P}\left(\boldsymbol{R}-\boldsymbol{R'}\right)\right]\nonumber \\
 & +\sum_{\boldsymbol{R'}}\phi_{P}\left(t,-\boldsymbol{R'}\right)\left[\Sigma^{HSEX}\left[\rho^{L}\right]\left(\boldsymbol{R}'\right)-\Sigma_{0}\left(\boldsymbol{R}'\right),\rho^{P}\left(\boldsymbol{R}-\boldsymbol{R'}\right)\right]\nonumber \\
 & +i\phi_{P}\left(t,-\boldsymbol{R}\right)\mathcal{L}_{incoh}\left[\rho^{L}\right]\left(\boldsymbol{R}\right),\label{eq:eom_peirls_gauge}
\end{align}
\end{widetext}where we can see that the term $\boldsymbol{E}\left(t\right).\boldsymbol{R}\rho^{L}\left(\boldsymbol{R}\right)$
is no longer involved in the propagation, as it is encoded in the
time-dependent Peierls phase, $\phi_{P}$. Eq. (\ref{eq:eom_peirls_gauge})
is what we call the semiconductor Wannier equations, a real-space
equivalent of the semiconductor Bloch equations. This is our preferred
formulation of the equations of motion. Up to this point, the problematic
transition dipole moments in the reciprocal space states did not appear
in any of our equations and therefore we avoid the problem of the
\emph{structure gauge} freedom. This is one of key features that makes
the SWEs more suitable than the SBEs

\subsubsection{Periodic boundary conditions}

The equations of motion for $\rho^{P}$, Eq. (\ref{eq:eom_peirls_gauge}),
are defined for an infinite crystal and for a practical implementation
one must truncate our sums over $\boldsymbol{R}$ to a finite set
of lattice vectors. The most common choice is to work in a supercell
approach. For that, we can define the supercell Bravais lattice as
\begin{equation}
\boldsymbol{R}_{S}=\sum^{D}_{i=1}d_{i}\boldsymbol{a}_{S,i},\ d_{i}\in\mathbb{Z},
\end{equation}
where $\boldsymbol{a}_{S,i}=N_{i}\boldsymbol{a}_{i}$ are the primitive
lattice vectors of the supercell, $N=\prod^{D}_{i=1}N_{i}$ is the
number of unit cells in the supercell. Furthermore, we impose Born--von
Karman periodic boundary conditions, i.e. $c^{\dagger}_{\boldsymbol{R}\alpha}\equiv c^{\dagger}_{\left(\boldsymbol{R}+\boldsymbol{R}_{S}\right)\alpha}$.
The Bravais lattice vectors enclosed into the origin supercell are
given by
\begin{equation}
\boldsymbol{R}=\sum^{D}_{i=1}n_{i}\boldsymbol{a}_{i},\ n_{i}\in\left[\lfloor-N_{i}/2\rfloor,\lfloor N_{i}/2-1\rfloor\right].\label{eq:supercell}
\end{equation}
From now on, the infinite sums in $\boldsymbol{R}$ will be replaced
by a sum of Bravais vectors in a single supercell, according to Eq.
(\ref{eq:supercell}). The choice of periodic boundary conditions
lead to a discretization of our Brillouin zone, i.e. 
\begin{equation}
\boldsymbol{k}=\sum^{D}_{i=1}\frac{m_{i}}{N_{i}}\boldsymbol{b}_{i},\ m_{i}\in\left[0,N_{i}-1\right],
\end{equation}
which defines a Monkhorst-Pack reciprocal space grid \citep{monkhorst1976special}.
In the following, all the sums over $\boldsymbol{k}$ will be assumed
to be done on this reciprocal space grid.

We can define the Bloch states associated with the localized Wannier
orbitals as 
\begin{align}
\left|\psi^{W}_{\boldsymbol{k}\alpha}\right\rangle  & =\frac{1}{\sqrt{N}}\sum_{\boldsymbol{R}}e^{i\boldsymbol{k}\cdot\boldsymbol{R}}\left|\boldsymbol{R}\alpha\right\rangle \\
c^{\dagger,W}_{\boldsymbol{k}\alpha} & =\frac{1}{\sqrt{N}}\sum_{\boldsymbol{R}}e^{i\boldsymbol{k}\cdot\boldsymbol{R}}c^{\dagger}_{\boldsymbol{R}\alpha}\\
c^{\dagger}_{\boldsymbol{R}\alpha} & =\frac{1}{\sqrt{N}}\sum_{\boldsymbol{k}}e^{-i\boldsymbol{k}\cdot\boldsymbol{R}}c^{\dagger,W}_{\boldsymbol{k}\alpha}
\end{align}
that are constructed from the localized Wannier orbitals. The upperscript
$W$ refers to the Wannier structure gauge \citep{wang2006ab,marzari2012maximally,silva2019high}.
It is important to distinguish between two distinct contexts in which
the term gauge is used: the \emph{laser gauge}, referring to the choice
between velocity and length gauges in the light-matter interaction,
and the \emph{structure gauge}, associated with the inherent freedom
in selecting a Bloch basis for the crystal Hamiltonian. In the $W$
gauge, we can also define the reduced density matrix in the reciprocal
space, $\rho^{W}_{\alpha\beta}\left(\boldsymbol{k}\right)=\left\langle c^{\dagger,W}_{\boldsymbol{k}\beta}c^{W}_{\boldsymbol{k}\alpha}\right\rangle $.
This RDM is related with $\rho^{L}\left(\boldsymbol{R}\right)$ by
a simple discrete Fourier transform 
\begin{align}
\rho^{W}\left(\boldsymbol{k}\right) & =\sum_{\boldsymbol{R}}e^{i\boldsymbol{k}\cdot\boldsymbol{R}}\rho^{L}\left(\boldsymbol{R}\right),\\
\rho^{L}\left(\boldsymbol{R}\right) & =\frac{1}{N}\sum_{\boldsymbol{k}}e^{-i\boldsymbol{k}\cdot\boldsymbol{R}}\rho^{W}\left(\boldsymbol{k}\right).
\end{align}
It is useful to define 
\begin{align}
\mathcal{F}^{L\rightarrow W}\left[\rho^{L}\left(\boldsymbol{R}\right)\right] & \equiv\rho^{W}\left(\boldsymbol{k}\right)\\
 & =\sum_{\boldsymbol{R}}e^{i\boldsymbol{k}\cdot\boldsymbol{R}}\rho^{L}\left(\boldsymbol{R}\right)
\end{align}
and the corresponding inverse transformation, $\mathcal{F}^{W\rightarrow L}$.

\subsubsection{Initial state}

In order to obtain the initial state, we must diagonalize $\hat{h}^{0}$.
In the basis, $\left|\psi^{W}_{\boldsymbol{k}\alpha}\right\rangle $,
$\hat{h}^{0}$ is expressed as
\begin{align}
\hat{h}^{0} & =\sum_{\boldsymbol{k}\alpha\beta}h^{0,W}_{\alpha\beta}\left(\boldsymbol{k}\right)c^{\dagger,W}_{\alpha\boldsymbol{k}}c^{W}_{\beta\boldsymbol{k}},\\
h^{0,W}\left(\boldsymbol{k}\right) & =\sum_{\boldsymbol{R}}e^{i\boldsymbol{k}.\boldsymbol{R}}h^{0}\left(\boldsymbol{R}\right)
\end{align}
and by diagonalizing $h^{0,W}\left(\boldsymbol{k}\right)$ we have
\begin{align}
\hat{h}^{0} & =\sum_{\boldsymbol{k}n}\varepsilon_{n}\left(\boldsymbol{k}\right)c^{\dagger,H}_{n\boldsymbol{k}}c^{H}_{n\boldsymbol{k}},\\
h^{0,H}_{nm}\left(\boldsymbol{k}\right) & =\delta_{nm}\varepsilon_{n}\left(\boldsymbol{k}\right)\\
h^{0,H}\left(\boldsymbol{k}\right)= & U^{\dagger}\left(\boldsymbol{k}\right)h^{0,W}\left(\boldsymbol{k}\right)U\left(\boldsymbol{k}\right),
\end{align}
where $U\left(\boldsymbol{k}\right)$ is the unitary matrix that diagonalizes
$h^{0,W}\left(\boldsymbol{k}\right)$, the upperscript $H$ refers
to the Hamiltonian structure gauge and $\varepsilon_{n}\left(\boldsymbol{k}\right)$
is the energy of band $n$. The equilibrium RDM in the Hamiltonian
gauge is just
\begin{equation}
\rho^{0,H}_{nm}\left(\boldsymbol{k}\right)=\delta_{nm}F_{\mu,T}\left(\varepsilon_{n}\left(\boldsymbol{k}\right)\right)
\end{equation}
where $F_{\mu,T}(\epsilon)=\left(e^{\beta(\epsilon-\mu)}+1\right)^{-1}$
is the Fermi-Dirac function for a chemical potential $\mu$ and inverse
temperature $\beta$. It is useful to define 
\begin{align}
\mathcal{U}^{W\rightarrow H}\left[\rho^{W}\left(\boldsymbol{k}\right)\right] & \equiv\rho^{H}\left(\boldsymbol{k}\right)\\
 & =U^{\dagger}\left(\boldsymbol{k}\right)\rho^{W}\left(\boldsymbol{k}\right)U\left(\boldsymbol{k}\right)
\end{align}
and the corresponding inverse transformation, $\mathcal{U}^{H\rightarrow W}$.
Note that the choice of $U\left(\boldsymbol{k}\right)$ is not unique.
This fact stems from the structure gauge freedom to choose the eigenstates,
i.e. $\left|\psi^{W}_{\boldsymbol{k}n}\right\rangle \rightarrow e^{i\theta_{n}\left(\boldsymbol{k}\right)}\left|\psi^{W}_{\boldsymbol{k}n}\right\rangle $.
However, one must notice that $\rho^{0,W}\left(\boldsymbol{k}\right)$
remains unaltered by this phase transformation. The $\mathcal{U}^{W\rightarrow H}$
transformation will allow to have access to populations of bands and
to compute terms that are inherent to the $H$ gauge, such as, interband
and intraband currents, and the inclusion of the pure dephasing term,
$\mathcal{L}_{D}$. 

\subsubsection{Observables}

The velocity operator, in contrast to the position operator, is invariant
under lattice translations. For a general invariant operator under
lattice translations, $\hat{O}$, its mean value normalized to the
unit cell can be expressed as
\begin{align}
\frac{\left\langle O\right\rangle }{N} & =\sum_{\alpha\beta\boldsymbol{R}}O_{\alpha\beta}\left(\boldsymbol{R}\right)\rho_{\beta\alpha}\left(-\boldsymbol{R}\right)\nonumber \\
 & =\sum_{\alpha\beta\boldsymbol{R}}O_{\alpha\beta}\left(\boldsymbol{R}\right)\left(\rho_{\alpha\beta}\left(\boldsymbol{R}\right)\right)^{*}.\label{eq:observable_crystal}
\end{align}
The explicit expression for the velocity operator matrix elements
in real space is 
\begin{equation}
\boldsymbol{v}\left(\boldsymbol{R}\right)=i\left(\boldsymbol{R}h^{0}\left(\boldsymbol{R}\right)-\sum_{\boldsymbol{R}'}\left[\boldsymbol{r}\left(\boldsymbol{R}'\right),h^{0}\left(\boldsymbol{R}-\boldsymbol{R}'\right)\right]\right).\label{eq:velocity_operator_crystal}
\end{equation}
We can recognize in the first (second) term of the velocity operator
the intraband (interband) term expressed in a real space basis \citep{silva2019high}.
We note that Berry-curvature effects are not neglected in the present
formulation. Although Eq. (\ref{eq:velocity_operator_crystal}) is
written in real space, its Fourier transform yields the standard multiband
velocity operator in reciprocal space \citep{silva2019high}. The
expression for the total current is, according to eq. (\ref{eq:observable_crystal}),
\begin{align}
\boldsymbol{j}\left(t\right) & =e\frac{\left\langle \boldsymbol{v}\right\rangle }{N}=e\sum_{\alpha\beta\boldsymbol{R}}\boldsymbol{v}_{\alpha\beta}\left(\boldsymbol{R}\right)\left(\rho_{\alpha\beta}\left(\boldsymbol{R}\right)\right)^{*}.
\end{align}
We must emphasize that observables must be calculated with $\rho^{L}$,
by perfoming the transformation defined in eq. (\ref{eq:lenght_peierls_transformation}),
as the corresponding operators do not take into account the Peierls
phase.

\subsubsection{Decoherence}

In Eq. (\ref{eq:eom_peirls_gauge}), we did not show the explicit
form of the incoherent terms that were grouped in $\mathcal{L}_{incoh}\left[\rho^{L}\right]$.
The first thing to notice is that we should move to the $L$ gauge.
In our work, as discussed in Sec. \ref{subsec:Decoherence}, we include
three forms of decoherence. The pure dephasing term, $\mathcal{L}_{D}$,
is naturally expressed in the $H$ gauge. By applying $\mathcal{F}^{L\rightarrow W}$
and $\mathcal{U}^{W\rightarrow H}$, one move $\rho$ to the $H$
gauge where $\mathcal{L}_{D}$ is simply 
\begin{equation}
\mathcal{L}_{D}\left[\rho^{H}\right]_{ij}\left(\boldsymbol{k}\right)=\gamma_{D}\left(\delta_{ij}-1\right)\rho^{H}_{ij}\left(\boldsymbol{k}\right)
\end{equation}
and after this apply the corresponding inverse transformations, $\mathcal{U}^{H\rightarrow W}$
and $\mathcal{F}^{W\rightarrow L}$.

The relaxation term, $\mathcal{L}_{r}$, is simply expressed as 
\begin{equation}
\mathcal{L}_{r}\left[\rho^{L}\right]\left(\boldsymbol{R}\right)=\gamma_{r}\left(\rho^{0,L}\left(\boldsymbol{R}\right)-\rho^{L}\left(\boldsymbol{R}\right)\right).
\end{equation}

Regarding the real space dephasing term, 
\begin{equation}
\mathcal{L}_{rs}\left[\rho^{L}\right]\left(\boldsymbol{R}\right)=\gamma_{rs}\left(d\left(\boldsymbol{R}\right)\right)\odot\left(\rho^{0,L}\left(\boldsymbol{R}\right)-\rho^{L}\left(\boldsymbol{R}\right)\right)
\end{equation}
where $d_{\alpha\beta}\left(\boldsymbol{R}\right)=\left|\boldsymbol{\tau}_{\alpha}-\boldsymbol{\tau}_{\beta}-\boldsymbol{R}\right|$,
$\boldsymbol{\tau}_{\alpha}$ is the center of the $\alpha$ Wannier
function in the center unit cell and $\odot$ denotes the Hadamard
product. The functional form of $\gamma_{rs}$ is given by Eq. (\ref{eq:rs_dephasing_form}).
Effectively, with this term we are suppressing coherences between
orbitals that are separated by distances greater than $R_{rs,cut}$
and and can therefore suppress trajectories that acquire a phase larger
than $2\pi$, and destructively interfere in the far-field \citep{Brown_2024,gaarde2008macroscopic}.
Having this analogy in mind, it will be useful to express $R_{rs,cut}$
in units of the excursion of an electron that acquires roughly a $2\pi$
phase, i.e. $R_{rs,0}=4\pi\omega_{L}/\left(\left|e\right|E_{0}\right)$,
where $E_{0}$ and $\omega_{L}$ are the amplitude and frequency of
the laser field.

\section{Results\label{sec:Results}}

In this section, we show numerical results that were obtained with
the SWEs formalism. We first show the behavior of the equilibrium
reduced density matrix and its decay. We also calculate the linear
optical conductivity for monolayer hBN and monolayer MoS$_{2}$, the
convergence of the SWEs with respect to the formulation of the SBEs
in \citep{silva2019high}, the impact of pure and real-space dephasing
in the HHG spectrum and finally the HHG spectrum for MoS$_{2}$.

\subsection{Decay of equilibrium density matrix}

\begin{figure}
\begin{centering}
\includegraphics[width=1\columnwidth]{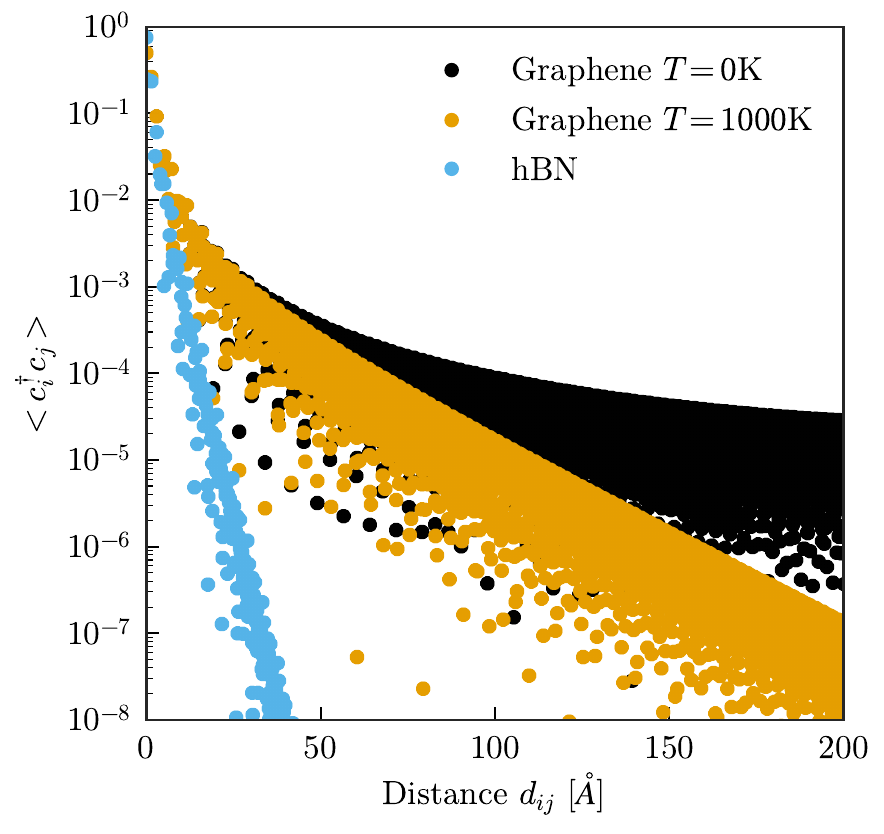}
\par\end{centering}
\caption{\label{fig:1}Scatter plot of the coherences between localized orbitals
for monolayer hBN (blue circles) and graphene, at zero (black circles)
and finite temperature (orange circles).}
\end{figure}

The basic object in the SWEs is the electronic reduced density matrix
expressed in a basis of Wannier localized orbitals. Despite working
with extended periodic systems, where Bloch states are delocalized,
the coherences of the electronic density matrix decay rapidly with
the distance between orbitals. \emph{Nearsightedness} of electronic
matter \citep{Kohn_1996,prodan2005nearsightedness} is actually the
core assumption for linear scaling electronic structure methods \citep{Goedecker_1999,prentice2020onetep}. 

In Fig. \ref{fig:1}, we show the decay of coherences in the density
matrix for graphene, at zero and finite temperature ($T=1000\,\mathrm{K}$),
and monolayer hBN. For both cases, we model graphene and monolayer
hBN with a tight-binding model \citep{castro2009electronic}, with
a nearest neighbor hopping $t_{0}=2.8\,\mathrm{eV}$ and a distance
between neighboring atoms of $1.4457\:\text{\AA}$. The gap energy
in the monolayer hBN is set to be $4.52\,\mathrm{eV}$. As expected,
hBN (a gapped insulator) and graphene at finite temperature (a finite
temperature semi-metal) exhibit an exponential decay of the coherences,
whereas graphene at zero temperature displays an algebraic decay \citep{Goedecker_1998,Goedecker_1999}.
It is the \emph{nearsightedness} of electrons, depicted in Fig. \ref{fig:1},
that make a real-space approach a compelling alternative to the reciprocal
space methods.

\subsection{Linear optical conductivity}

\begin{figure}
\begin{centering}
\includegraphics[width=1\columnwidth]{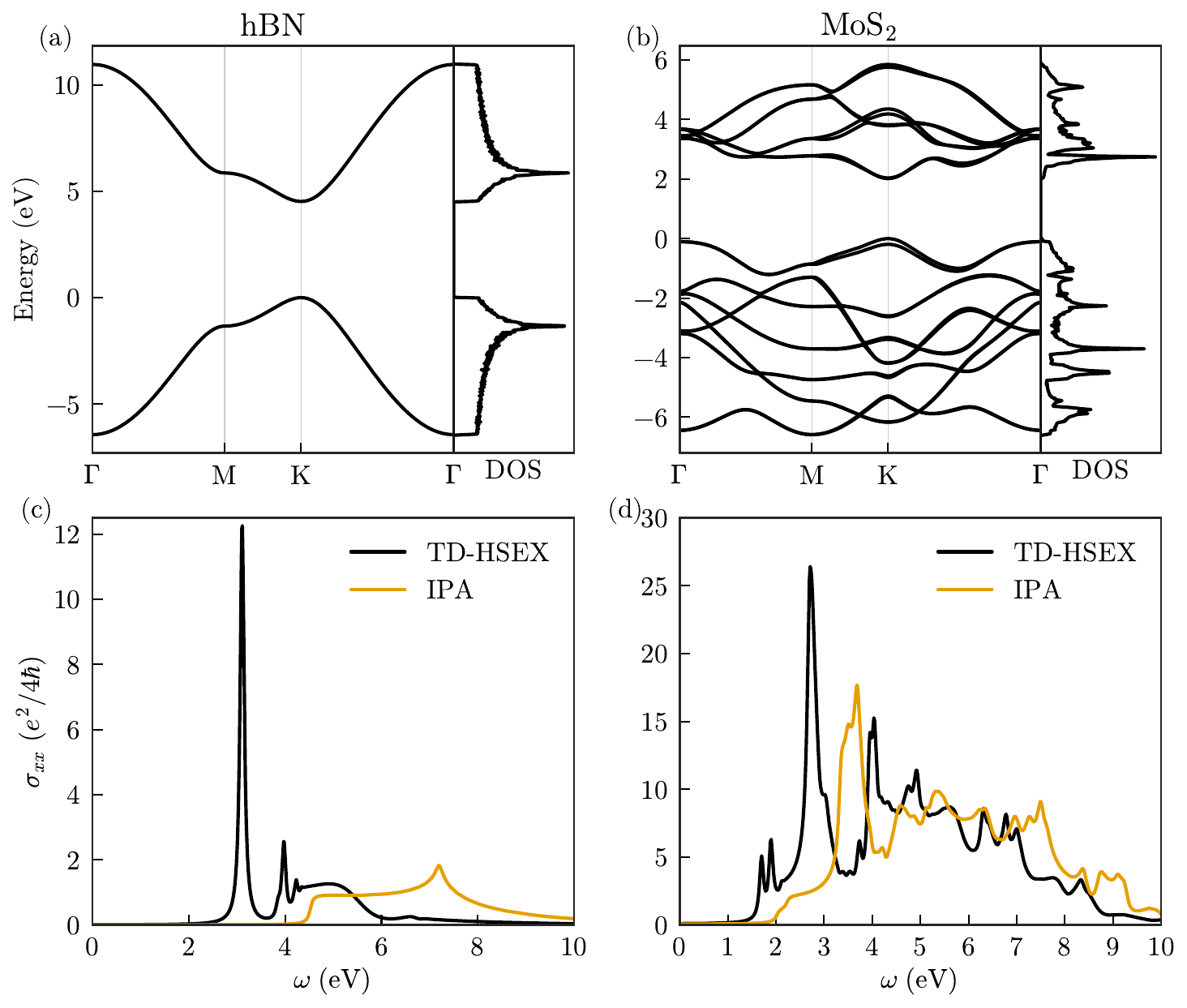}
\par\end{centering}
\caption{\label{fig:2}Band-structure and total density of states (DOS) of
monolayer hBN (a) and of monolayer 1H-MoS$_{2}$ (b). In panels (c,d),
we show the linear optical conductivity of monolayer hBN and 1H-MoS$_{2}$,
respectively, in the IPA (orange lines) and TD-HSEX (black lines)
approximations.}
\end{figure}

To validate the SWEs approach, we have calculated the linear optical
conductivity of hBN and MoS$_{2}$. In both cases we have a point
group symmetry $D_{3h}$ and for symmetry reasons, $\sigma_{xy}=0$
and $\sigma_{xx}=\sigma_{yy}$ \citep{ventura2021nonlinear}. We calculate
the linear optical conductivity by applying a very short pulse, with
an field strength in the linear response regime, $E_{0}=2\,\mathrm{kV/m}$,
and a FWHM in electric field of $0.157\,\mathrm{fs}$ for a $\cos^{2}$
envelope. The linear optical conductivity can be obtained by using
Ohm's law, $\sigma\left(\omega\right)=\tilde{j}\left(\omega\right)/\tilde{E}\left(\omega\right)A_{UC}$,
where $\tilde{j}\left(\omega\right)=\int^{+\infty}_{-\infty}dt\exp\left(-i\omega t\right)j\left(t\right)$
and $A_{UC}$ is the unit-cell area. To dampen the signal, we introduce
a relaxation parameter of $\hbar\gamma_{r}=0.1\,\mathrm{eV}$. To
model the screening potential, we used the Rytova-Keldysh potential
(see Appendix \ref{sec:Rytova-Keldysh-potential}), and used for the
monolayer hBN (MoS$_{2}$) a screening length $r_{0}=10\,\text{\AA}$
($r_{0}=13.55\,\text{\AA}$) and set the dielectric constants as $\left(\epsilon_{1}+\epsilon_{2}\right)/2=1$
($\left(\epsilon_{1}+\epsilon_{2}\right)/2=2.5$). The Rytova--Keldysh
screening potential used for both hBN and MoS$_{2}$ is taken from
the literature \citep{ridolfi2020expeditious}. In both cases, we
used a supercell with $N_{1}=N_{2}=192$. The bandstructure of MoS$_{2}$
was obtained by performing an \emph{ab-initio} calculation using the
HSE06 functional \citep{heyd2003hybrid}, including spin-orbital coupling,
in a 16$\times$16 Monkhorst-Pack grid using the QuantumEspresso code
\citep{QE-2017}. We perform a projection on the $p$ orbitals of
S and $d$ orbitals of Mo, and a Wannierization procedure to obtain
the Hamiltonian and dipole couplings using the Wannier90 software
\citep{mostofi2014updated}, obtaining 22 bands to model MoS$_{2}$. 

In Fig. \ref{fig:2}, we show in panels (a,b) the bandstructure and
density of states (DOS) for hBN and MoS$_{2}$. In panels (c,d), we
plot the linear optical conductivity for hBN and MoS$_{2}$ in the
independent-particle approximation (IPA), i.e. setting $\Sigma^{HSEX}=0$,
and in the TD-HSEX approximation. First, our results match very well
previous studies \citep{ridolfi2020expeditious,cistaro2022theoretical},
both in the IPA and TD-HSEX approximations. Furthermore, the TD-HSEX
is able to capture correctly optical excitons in hBN and MoS$_{2}$.

\subsection{Convergence: SWEs vs. SBEs}

\begin{figure}
\begin{centering}
\includegraphics[width=1\columnwidth]{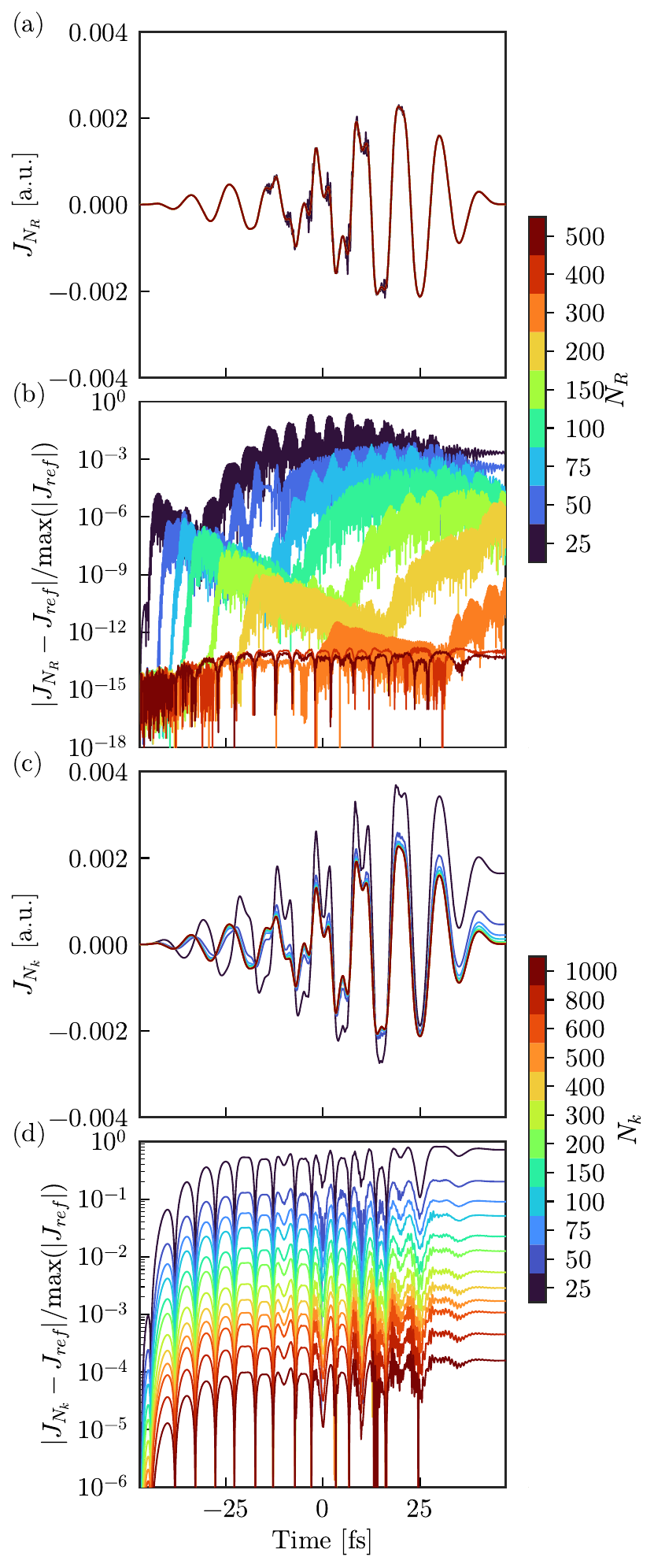}
\par\end{centering}
\caption{\label{fig:3}Calculated current (a) and the relative error of the
current (b) using the SWEs approach for different $N_{R}$. (c,d)
The same for calculations using the SBEs approach.}
\end{figure}

We benchmarked the numerical convergence of the SBEs in momentum space,
using the formalism of \citep{silva2019high}, against that of the
SWEs in real space by calculating the current generated in monolayer
hBN under illumination by a mid-infrared laser pulse. The laser field
is modeled as
\begin{equation}
\boldsymbol{E}\left(t\right)=E_{0}f\left(t\right)\sin\left(\omega t+\phi\right)\hat{\boldsymbol{\mathrm{e}}}
\end{equation}
where $E_{0}=40\,\mathrm{MV/cm}$ polarized in the $\Gamma-\mathrm{M}$
direction (along the N-B bond), with a central wavelength of $3\,\mu\mathrm{m}$
and a carrier envelope phase $\phi=\pi/2$. The pulse envelope $f\left(t\right)$
is taken to be a $\cos^{2}$ profile with a FWHM in electric field
of $47.1\,\mathrm{fs}$. We introduce in the calculations pure dephasing
with $T_{2}=5\,\mathrm{fs}$. For both cases, we integrate the equations
of motion using a Runge-Kutta 4 propagator with a timestep of $dt=0.1\,\mathrm{a.u.}$.

In Fig. \ref{fig:3}, we show the calculated current in the SWEs formalism
(a,b) and in the SBEs formalism (c,d). For the SWEs, we calculate
the current using different supercells, where $N_{1}=N_{2}=N_{R}$,
and we take as reference a calculation with $N_{R}=600$. In the SBEs,
we discretize the reciprocal space in a $N_{k}\times N_{k}$ Monkhorst-Pack
grid and we take as reference a calculation with $N_{k}=1200$. We
can see in Fig. \ref{fig:3}(b) that the relative error decreases
quite fast with a moderate $N_{R}$, noticing that with $N_{R}=400$
we reach numerical precision accuracy. On the other hand, the SBEs
converge much slower when compared with the SWEs, see Fig. \ref{fig:3}(b,d).
To achieve a relative error of around $10^{-3}$, a supercell with
$N_{R}=50$ is sufficient. While in the SBEs case, we need to increase
the $k$-space grid to $N_{k}=600$. The slower convergence of the
SBEs calculations originates from the finite-difference discretization
of the $k$-derivative \citet{silva2019high}, which is avoided in
the SWEs formulation. One additional advantage of the SWEs over the
SBEs approach is numerical: convergence is achieved significantly
faster, enabling the exploration of more complex materials that would
otherwise be computationally prohibitive within the SBEs framework.

\subsection{Real-space dephasing }

\begin{figure}
\begin{centering}
\includegraphics[width=1\columnwidth]{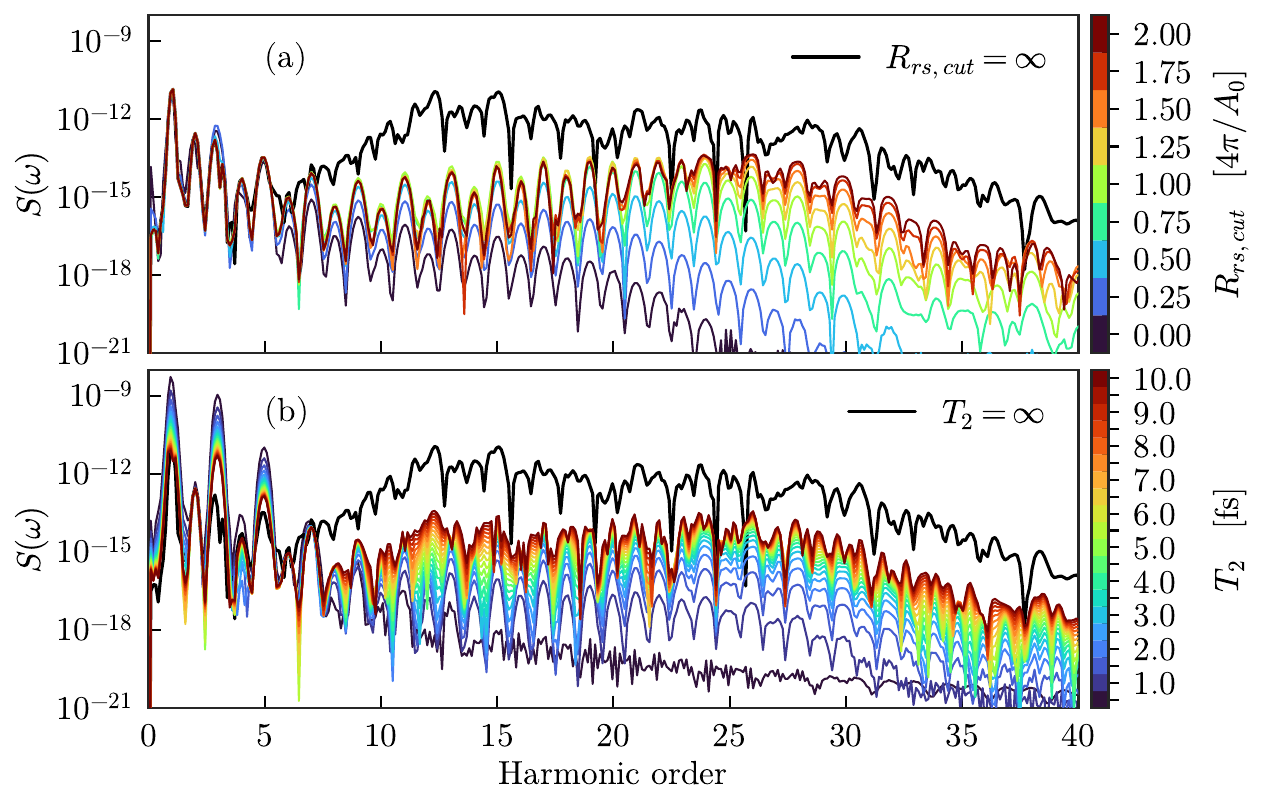}
\par\end{centering}
\caption{\label{fig:4}High harmonic spectrum of monolayer hBN with different
dephasing mechanisms: (a) real-space dephasing and (b) pure dephasing.}
\end{figure}

The requirement for ultrashort dephasing times to match experimental
high-harmonic spectra in solids remains an open question. A recently
proposed method seeks to address this issue by introducing real-space
dephasing \citep{Brown_2024}, which arises naturally within the SWEs
framework.

Fig. \ref{fig:4} compares the two approaches by calculating the high-harmonic
spectrum in hBN while scanning $R_{rs,cut}$ for the real-space dephasing
case and $T_{2}$ for the pure-dephasing case. For real-space dephasing,
we set $\alpha_{rs}=2$ and $\beta_{rs}=6.7736\times10^{-5}\,\mathrm{a.u.}$.
The laser field is the same as in Fig. \ref{fig:3}. In these calculations,
we used a supercell with $N_{1}=N_{2}=100$ and propagate the SWEs
with timestep $dt=0.5\,\mathrm{a.u.}$ using a Dormand-Prince's 5/4
Runge-Kutta method. The harmonic spectrum is calculated as the square
of the Fourier transform of the dipole acceleration \citep{yue2022introduction},
$\boldsymbol{a}\left(t\right)=\frac{d\boldsymbol{j}\left(t\right)}{dt}$.
To avoid artificial spectral features caused by abrupt discontinuities
at the end of the pulse, a window function is applied to the current
prior to the Fourier transform \citep{silva2016thesis}.

Both real-space and pure dephasing reduce interferences in the high-harmonic
signal. However, real-space dephasing preserves the intensity of low-order
harmonics, matching the purely coherent case. In contrast, pure dephasing
strongly enhances low-order harmonics, an artifact of the extremely
short dephasing times, primarily caused by dephasing-induced ionization
\citep{boroumand2025strong}. Conventional $T_{2}$-type dephasing
enhances the conduction-band population through dephasing-induced
excitation, thereby increasing the low-order harmonics, whereas real-space
dephasing largely preserves the short trajectories responsible for
this spectral region \citep{molinero2025phd}. For these reasons,
the real-space dephasing approach is a more reliable and physically
consistent alternative to simple pure-dephasing models.

\subsection{HHG in MoS$_{2}$}

\begin{figure*}
\begin{centering}
\includegraphics[width=2\columnwidth]{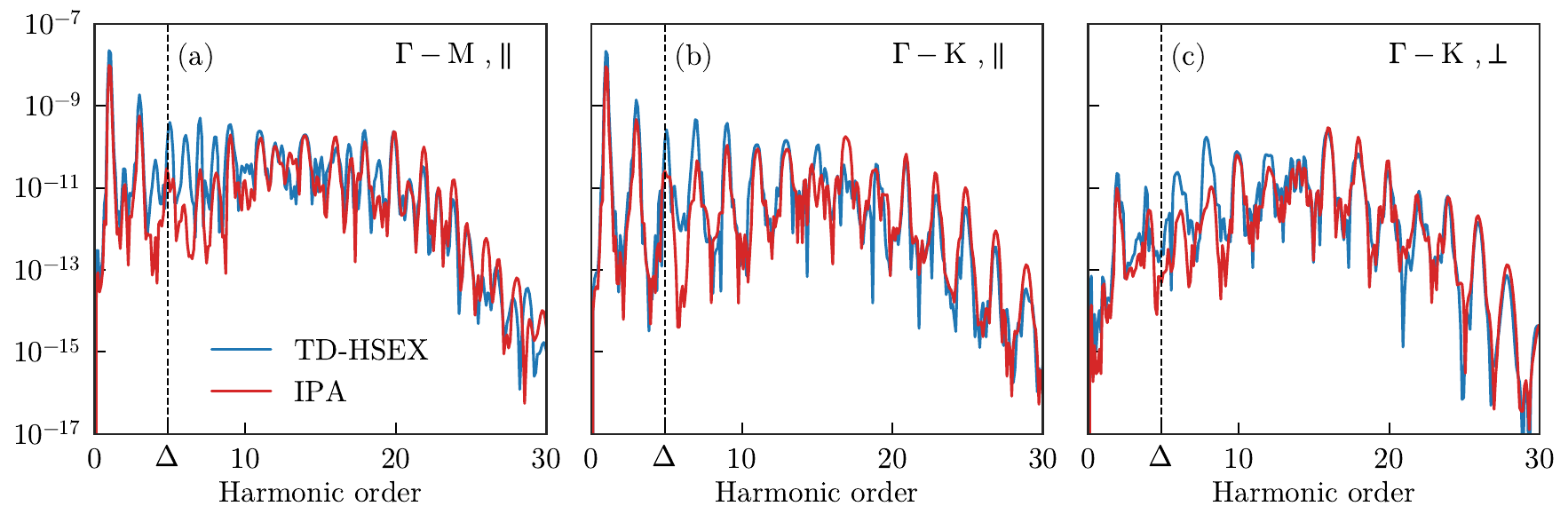}
\par\end{centering}
\caption{\label{fig:5}High harmonic spectrum at the IPA and TD-HSEX level
in MoS$_{2}$. (a) HHG spectrum for a laser polarized along the $\Gamma-\mathrm{M}$
direction in the parallel direction. (b,c) Same as in (a) for a laser
along the $\Gamma-\mathrm{K}$ in the parallel and perpendicular direction,
respectively. The dashed vertical line corresponds to the bandgap
of MoS$_{2}$. }
\end{figure*}

We have calculated the HHG spectra in MoS$_{2}$. All parameters of
the laser pulse are the same as in Fig. \ref{fig:3}, apart from the
peak field, which is $E_{0}=30\,\mathrm{MV/cm}$.We introduce a pure-dephasing
term, $T_{2}=4\,\mathrm{fs}$. In these calculations, we used a supercell
with $N_{1}=N_{2}=100$ and propagate the SWEs with timestep $dt=0.5\,\mathrm{a.u.}$
using a Dormand-Prince's 5/4 Runge-Kutta method.

Fig. \ref{fig:5} shows the harmonic spectra for different laser polarization
and emission directions. Symmetry requires that for a laser pulse
oriented along the $\Gamma-\mathrm{M}$ direction, the perpendicular
current vanishes and is therefore not displayed. In contrast, for
emission parallel to the pulse in this direction, both even and odd
harmonics are allowed because inversion symmetry is broken. When the
laser pulse is aligned along the $\Gamma-\mathrm{K}$ direction, symmetry
constrains the harmonic emission: only odd (even) harmonics appear
for parallel (perpendicular) emission \citep{you2017anisotropic}.
We observe that when excitonic effects are taken into account (TD-HSEX),
we see an enhancement of the harmonic signal in harmonics below the
bandgap ($\Delta$), that is consistent with previous findings in
the literature \citep{molinero2024subcycle}.

\section{Conclusions\label{sec:conclusions}}

We have introduced the semiconductor Wannier equations (SWEs), a real-time,
real-space framework for describing ultrafast light--matter interaction
and nonlinear optical response in crystalline solids. By expressing
the electronic reduced density matrix in a localized Wannier basis,
the SWEs provide a gauge-clean alternative to reciprocal-space semiconductor
Bloch equations (SBEs), avoiding the numerical instabilities caused
by the structure-gauge freedom of Bloch states. The method naturally
incorporates electron--electron interactions at the time-dependent
Hartree plus screened-exchange (TD-HSEX) level and includes physically
motivated decoherence mechanisms, pure dephasing, population relaxation,
and distance-dependent real-space dephasing, offering a robust modeling
of strong-field and high-harmonic generation dynamics.

We have shown that the SWEs reproduce key optical properties, such
as the linear optical conductivity and excitonic features of monolayer
hBN and MoS$_{2}$, in agreement with established approaches. Benchmarking
against the SBEs formalism demonstrated that the SWEs achieve significantly
faster numerical convergence, enabling efficient simulations with
moderate supercell sizes. Importantly, the real-space dephasing model
introduced here provides a physically grounded alternative to the
ultrashort pure-dephasing times often used to fit experimental HHG
spectra, avoiding artificial enhancement of low-order harmonics while
preserving the coherent features of the signal.

Conceptually, the SWEs bridge semiclassical real-space intuition,
central to attosecond physics, with many-body solid-state optics,
opening new avenues for interpreting HHG and other strong-field phenomena
in complex materials. Their computational efficiency, robustness to
gauge ambiguities, and compatibility with \emph{ab-initio} Hamiltonians
make the SWEs a promising platform for exploring nonlinear optical
response, excitonic dynamics, and attosecond spectroscopy in emerging
quantum materials.

Future developments may include the incorporation of fully dynamical
screening, coupling to phonons and other bosonic modes, and extensions
to strongly correlated systems beyond mean-field approximations. These
advances will further strengthen the SWEs as a versatile and predictive
tool for next-generation ultrafast solid-state optics.

\section*{Acknowledgments}

R.E.F.S. acknowledge fruitful discussions with Johannes Feist. This
work was supported through Grants PID2021-122769NB-I00, RYC2022-035373-I
and CNS2024-154463, funded by MICIU/AEI/10.13039/501100011033, “ERDF
A way of making Europe” and “ESF+”. A.J.G. acknowledges support from
the Talento Comunidad de Madrid Fellowship 2022-T1/IND24102 and the
Spanish Ministry of Science, Innovation and Universities through grant
reference PID2023-146676NA-I00. B.A. and J.M.V.P.L. acknowledge support
by the Portuguese Foundation for Science and Technology (FCT) in the
framework of the Strategic Funding UID/04650/2025.

\appendix

\section{Rytova-Keldysh potential\label{sec:Rytova-Keldysh-potential}}

As we are dealing with 2D materials, the Rytova-Keldysh potential
\citep{rytova2018screened,keldysh1979coulomb,Van_Tuan_2018} is our
choice to model screening effects in the self-energy term, $\Sigma^{SEX}$.
The Rytova-Keldysh potential reads 
\begin{equation}
W_{RK}\left(r\right)=\frac{e^{2}}{4\epsilon_{0}\left(\epsilon_{1}+\epsilon_{2}\right)r_{0}}\left[H_{0}\left(\frac{r}{r_{0}}\right)-Y_{0}\left(\frac{r}{r_{0}}\right)\right],
\end{equation}
where $\epsilon_{1}$ and $\epsilon_{2}$ are the dielectric constants
of the top and bottom medium, $r_{0}$ is the screening length and
$H_{0}$, $Y_{0}$ are the zero-order Struve and Neumann special functions.
In order to avoid the divergence of the potential at $r=0$ \citep{Wu_2015,uria2024efficient},
we renormalize $r\rightarrow\sqrt{r^{2}+r^{2}_{min}}$. Furthermore,
for large distances, i.e. $r>r_{RK,cut}$, we employ a radial cutoff
and take $W_{RK}$ to be zero. In pratice, to avoid artifacts from
periodic images in the calculation, we will set $r_{RK,cut}$ to be
of the radius of the largest sphere that fits inside the supercell.
The renormalization and radial cutoff are also applied to the bare
Coulomb potential in the Hartree term, $\Sigma^{H}$.

\bibliography{atata_paper}

@article{molinero2024subcycle,
  title={Subcycle dynamics of excitons under strong laser fields},
  author={Molinero, Eduardo B and Amorim, Bruno and Malakhov, Mikhail and Cistaro, Giovanni and Jim{\'e}nez-Gal{\'a}n, {\'A}lvaro and Pic{\'o}n, Antonio and San-Jos{\'e}, Pablo and Ivanov, Misha and Silva, Rui EF},
  journal={Science advances},
  volume={10},
  number={35},
  pages={eadn6985},
  year={2024},
  publisher={American Association for the Advancement of Science}
}

@article{you2017anisotropic,
  title={Anisotropic high-harmonic generation in bulk crystals},
  author={You, Yong Sing and Reis, David A and Ghimire, Shambhu},
  journal={Nature physics},
  volume={13},
  number={4},
  pages={345--349},
  year={2017},
  publisher={Nature Publishing Group UK London}
}

@article{QE-2017,
	author = {Giannozzi, P and Andreussi, O and Brumme, T and Bunau, O and Nardelli, M Buongiorno and Calandra, M and Car, R and Cavazzoni, C and Ceresoli, D and Cococcioni, M and Colonna, N and Carnimeo, I and Corso, A Dal and de Gironcoli, S and Delugas, P and DiStasio, Jr, R A and Ferretti, A and Floris, A and Fratesi, G and Fugallo, G and Gebauer, R and Gerstmann, U and Giustino, F and Gorni, T and Jia, J and Kawamura, M and Ko, H-Y and Kokalj, A and K{\"u}{\c{c}}{\"u}kbenli, E and Lazzeri, M and Marsili, M and Marzari, N and Mauri, F and Nguyen, N L and Nguyen, H-V and Otero-de-la-Roza, A and Paulatto, L and Ponc{\'e}, S and Rocca, D and Sabatini, R and Santra, B and Schlipf, M and Seitsonen, A P and Smogunov, A and Timrov, I and Thonhauser, T and Umari, P and Vast, N and Wu, X and Baroni, S},
	doi = {10.1088/1361-648X/aa8f79},
	journal = {Journal of Physics: Condensed Matter},
	number = {46},
	pages = {465901},
	title = {Advanced capabilities for materials modelling with QUANTUM ESPRESSO},
	url = {http://stacks.iop.org/0953-8984/29/i=46/a=465901},
	volume = {29},
	year = {2017}
}

@article{mostofi2014updated,
	author = {Mostofi, Arash A and Yates, Jonathan R and Pizzi, Giovanni and Lee, Young-Su and Souza, Ivo and Vanderbilt, David and Marzari, Nicola},
	doi = {10.1016/j.cpc.2014.05.003},
	journal = {Computer Physics Communications},
	number = {8},
	pages = {2309--2310},
	publisher = {Elsevier},
	title = {An updated version of wannier90: A tool for obtaining maximally-localised Wannier functions},
	volume = {185},
	year = {2014}
}

@article{Sanchez-Barquilla2020,
	author = {{S{\'a}nchez-Barquilla}, M. and Silva, R. E. F. and Feist, J.},
	copyright = {All rights reserved},
	doi = {10.1063/1.5138937},
	issn = {0021-9606},
	journal = {J. Chem. Phys.},
	number = {3},
	pages = {034108},
	title = {Cumulant Expansion for the Treatment of Light-Matter Interactions in Arbitrary Material Structures},
	volume = {152},
	year = {2020}
}

@misc{QuantumAlgebra.jl,
	author = {Feist, Johannes and contributors},
	doi = {10.5281/zenodo.3525845},
	month = {9},
	title = {QuantumAlgebra.jl},
	url = {https://github.com/jfeist/QuantumAlgebra.jl},
	version = {v1.1.0},
	year = {2021}
}

@article{Kira_2008,
	author = {Kira, M. and Koch, S. W.},
	doi = {10.1103/physreva.78.022102},
	issn = {1094-1622},
	journal = {Physical Review A},
	number = {2},
	publisher = {American Physical Society (APS)},
	title = {Cluster-expansion representation in quantum optics},
	volume = {78},
	year = {2008}
}

@book{Kira_2011,
	author = {Kira, Mackillo and Koch, Stephan W.},
	doi = {10.1017/cbo9781139016926},
	isbn = {9781139016926},
	publisher = {Cambridge University Press},
	title = {Semiconductor Quantum Optics},
	year = {2011}
}

@book{bruus2004many,
	author = {Bruus, Henrik and Flensberg, Karsten},
	publisher = {Oxford university press},
	title = {Many-body quantum theory in condensed matter physics: an introduction},
	year = {2004}
}

@mastersthesis{ribeiro2023excitonic,
	author = {Ribeiro, Francisco Ricardo Lobo},
	school = {Universidade do Minho (Portugal)},
	title = {Excitonic properties of hBN from a time-dependent hartree-fock mean-field theory},
	url = {https://hdl.handle.net/1822/91615},
	year = {2023}
}

@book{martinbookinteracting,
	author = {Martin, R. M. and Reining, L. and Ceperley, D. M.},
	publisher = {Cambridge University Press},
	title = {Interacting Electrons: Theory and Computational Approaches},
	year = {2016}
}

@book{martin2004electronic,
	author = {Martin, Richard},
	journal = {Pr., West Nyack, NY},
	publisher = {Cambridge University Press},
	title = {Electronic Structure: Basic Theory and Practical Methods},
	year = {2020}
}

@article{Attaccalite_2011,
	author = {Attaccalite, C. and Gr{\"o}ning, M. and Marini, A.},
	doi = {10.1103/physrevb.84.245110},
	issn = {1550-235X},
	journal = {Physical Review B},
	number = {24},
	publisher = {American Physical Society (APS)},
	title = {Real-time approach to the optical properties of solids and nanostructures: Time-dependent Bethe-Salpeter equation},
	volume = {84},
	year = {2011}
}

@article{sangalli2021excitons,
	author = {Sangalli, D},
	doi = {10.1103/PhysRevMaterials.5.083803},
	journal = {Physical Review Materials},
	number = {8},
	pages = {083803},
	publisher = {APS},
	title = {Excitons and carriers in transient absorption and time-resolved ARPES spectroscopy: An ab initio approach},
	volume = {5},
	year = {2021}
}

@article{sangalli2018ab,
	author = {Sangalli, Davide and Perfetto, Enrico and Stefanucci, Gianluca and Marini, Andrea},
	doi = {10.1140/epjb/e2018-90126-5},
	journal = {The European Physical Journal B},
	number = {8},
	pages = {171},
	publisher = {Springer},
	title = {An ab-initio approach to describe coherent and non-coherent exciton dynamics},
	volume = {91},
	year = {2018}
}

@article{sangalli2023exciton,
	author = {Sangalli, D and d'Alessandro, M and Attaccalite, Claudio},
	doi = {10.1103/PhysRevB.107.205203},
	journal = {Physical Review B},
	number = {20},
	pages = {205203},
	publisher = {APS},
	title = {Exciton-exciton transitions involving strongly bound excitons: an ab initio approach},
	volume = {107},
	year = {2023}
}

@article{Brown_2024,
	author = {Brown, Graham G. and Jim{\'e}nez-Gal{\'a}n, {\'A}lvaro and Silva, Rui E. F. and Ivanov, Misha},
	doi = {10.1103/physrevresearch.6.043005},
	journal = {Physical Review Research},
	number = {4},
	publisher = {American Physical Society (APS)},
	title = {Real-space perspective on dephasing in solid-state high harmonic generation},
	volume = {6},
	year = {2024}
}

@article{Brown_2024josa,
	author = {Brown, Graham G. and Jim{\'e}nez-Gal{\'a}n, {\'a}lvaro and Silva, Rui E. F. and Ivanov, Misha},
	doi = {10.1364/josab.513543},
	issn = {1520-8540},
	journal = {Journal of the Optical Society of America B},
	number = {6},
	pages = {B40},
	publisher = {Optica Publishing Group},
	title = {Ultrafast dephasing in solid-state high harmonic generation: macroscopic origin revealed by real-space dynamics [Invited]},
	volume = {41},
	year = {2024}
}

@article{Floss_2018,
	author = {Floss, Isabella and Lemell, Christoph and Wachter, Georg and Smejkal, Valerie and Sato, Shunsuke A. and Tong, Xiao-Min and Yabana, Kazuhiro and Burgd{\"o}rfer, Joachim},
	doi = {10.1103/physreva.97.011401},
	issn = {2469-9934},
	journal = {Physical Review A},
	number = {1},
	publisher = {American Physical Society (APS)},
	title = {Ab initio multiscale simulation of high-order harmonic generation in solids},
	volume = {97},
	year = {2018}
}

@article{Vampa_2014,
	author = {Vampa, G. and McDonald, C. R. and Orlando, G. and Klug, D. D. and Corkum, P. B. and Brabec, T.},
	doi = {10.1103/physrevlett.113.073901},
	issn = {1079-7114},
	journal = {Physical Review Letters},
	number = {7},
	publisher = {American Physical Society (APS)},
	title = {Theoretical Analysis of High-Harmonic Generation in Solids},
	volume = {113},
	year = {2014}
}

@article{terada2024limitations,
	author = {Terada, Ibuki and Kitamura, Sota and Watanabe, Hiroshi and Ikeda, Hiroaki},
	doi = {10.1103/PhysRevB.109.L180302},
	journal = {Physical Review B},
	number = {18},
	pages = {L180302},
	publisher = {APS},
	title = {Limitations and improvements of the relaxation time approximation in the quantum master equation: Linear conductivity in insulating systems},
	volume = {109},
	year = {2024}
}

@article{cox2014electrically,
	author = {Cox, Joel D and {Javier Garc{\'i}a de Abajo}, F},
	doi = {10.1038/ncomms6725},
	journal = {Nature communications},
	number = {1},
	pages = {5725},
	publisher = {Nature Publishing Group UK London},
	title = {Electrically tunable nonlinear plasmonics in graphene nanoislands},
	volume = {5},
	year = {2014}
}

@article{sato2021nonlinear,
	author = {Sato, Shunsuke A and Rubio, Angel},
	doi = {10.1088/1367-2630/ac03d0},
	journal = {New Journal of Physics},
	number = {6},
	pages = {063047},
	publisher = {IOP Publishing},
	title = {Nonlinear electric conductivity and THz-induced charge transport in graphene},
	volume = {23},
	year = {2021}
}

@article{silva2016even,
	author = {Silva, REF and Rivi{\`e}re, Paula and Morales, F and Smirnova, O and Ivanov, M and Mart{\'i}n, Fernando},
	doi = {10.1038/srep32653},
	journal = {Scientific reports},
	number = {1},
	pages = {32653},
	publisher = {Nature Publishing Group UK London},
	title = {Even harmonic generation in isotropic media of dissociating homonuclear molecules},
	volume = {6},
	year = {2016}
}

@article{marzari2012maximally,
	author = {Marzari, Nicola and Mostofi, Arash A and Yates, Jonathan R and Souza, Ivo and Vanderbilt, David},
	doi = {10.1103/RevModPhys.84.1419},
	journal = {Reviews of Modern Physics},
	number = {4},
	pages = {1419--1475},
	publisher = {APS},
	title = {Maximally localized Wannier functions: Theory and applications},
	volume = {84},
	year = {2012}
}

@incollection{blount1962formalisms,
	author = {Blount, Eugene I},
	booktitle = {Solid state physics},
	doi = {10.1016/S0081-1947(08)60459-2},
	pages = {305--373},
	publisher = {Elsevier},
	title = {Formalisms of band theory},
	volume = {13},
	year = {1962}
}

@phdthesis{ventura2021nonlinear,
	author = {Ventura, Gon{\c{c}}alo Bastos},
	school = {Universidade do Porto (Portugal)},
	title = {Nonlinear optical response of two-dimensional crystals},
	url = {https://www.proquest.com/docview/2926441819?pq-origsite=gscholar&fromopenview=true&sourcetype=Dissertations%20&%20Theses},
	year = {2021}
}

@phdthesis{silva2016thesis,
	author = {Silva, R E F},
	school = {Universidad Autonoma de Madrid (Spain)},
	title = {Study of diatomic molecules under short intense laser pulses},
	url = {https://arxiv.org/abs/1705.06590},
	year = {2016}
}

@article{ventura2017gauge,
	author = {Ventura, GB and Passos, DJ and {Lopes dos Santos}, JMB and {Viana Parente Lopes}, JM and Peres, Nuno MR},
	doi = {10.1103/PhysRevB.96.035431},
	journal = {Physical Review B},
	number = {3},
	pages = {035431},
	publisher = {APS},
	title = {Gauge covariances and nonlinear optical responses},
	volume = {96},
	year = {2017}
}

@article{silva2019high,
	author = {Silva, REF and Mart{\'i}n, Fernando and Ivanov, M},
	doi = {10.1103/PhysRevB.100.195201},
	journal = {Physical Review B},
	number = {19},
	pages = {195201},
	publisher = {APS},
	title = {High harmonic generation in crystals using maximally localized Wannier functions},
	volume = {100},
	year = {2019}
}

@article{cistaro2022theoretical,
	author = {Cistaro, Giovanni and Malakhov, Mikhail and Esteve-Paredes, Juan Jos{\'e} and Ur{\'i}a-{\'A}lvarez, Alejandro Jos{\'e} and Silva, Rui EF and Mart{\'i}n, Fernando and Palacios, Juan Jos{\'e} and Pic{\'o}n, Antonio},
	doi = {10.1021/acs.jctc.2c00674},
	journal = {Journal of Chemical Theory and Computation},
	number = {1},
	pages = {333--348},
	publisher = {ACS Publications},
	title = {Theoretical approach for electron dynamics and ultrafast spectroscopy (EDUS)},
	volume = {19},
	year = {2022}
}

@article{marzari1997maximally,
	author = {Marzari, Nicola and Vanderbilt, David},
	doi = {10.1103/PhysRevB.56.12847},
	journal = {Physical review B},
	number = {20},
	pages = {12847},
	publisher = {APS},
	title = {Maximally localized generalized Wannier functions for composite energy bands},
	volume = {56},
	year = {1997}
}

@article{gaarde2008macroscopic,
	author = {Gaarde, Mette B and Tate, Jennifer L and Schafer, Kenneth J},
	doi = {10.1088/0953-4075/41/13/132001},
	journal = {Journal of Physics B: Atomic, Molecular and Optical Physics},
	number = {13},
	pages = {132001},
	publisher = {IOP Publishing},
	title = {Macroscopic aspects of attosecond pulse generation},
	volume = {41},
	year = {2008}
}

@article{scrinzi2010infinite,
	author = {Scrinzi, Armin},
	doi = {10.1103/PhysRevA.81.053845},
	journal = {Physical Review A},
	number = {5},
	pages = {053845},
	publisher = {APS},
	title = {Infinite-range exterior complex scaling as a perfect absorber in time-dependent problems},
	volume = {81},
	year = {2010}
}

@article{pedatzur2015attosecond,
	author = {Pedatzur, Oren and Orenstein, Gal and Serbinenko, V and Soifer, Hadas and Bruner, BD and Uzan, AJ and Brambila, DS and Harvey, AG and Torlina, L and Morales, F and others},
	doi = {10.1038/nphys3436},
	journal = {Nature Physics},
	number = {10},
	pages = {815--819},
	publisher = {Nature Publishing Group UK London},
	title = {Attosecond tunnelling interferometry},
	volume = {11},
	year = {2015}
}

@article{richter2013role,
	author = {Richter, Maria and Patchkovskii, Serguei and Morales, Felipe and Smirnova, Olga and Ivanov, Misha},
	doi = {10.1088/1367-2630/15/8/083012},
	journal = {New Journal of Physics},
	number = {8},
	pages = {083012},
	publisher = {IOP Publishing},
	title = {The role of the Kramers--Henneberger atom in the higher-order Kerr effect},
	volume = {15},
	year = {2013}
}

@article{yue2022introduction,
	author = {Yue, Lun and Gaarde, Mette B},
	doi = {10.1364/JOSAB.448602},
	journal = {Journal of the Optical Society of America B},
	number = {2},
	pages = {535--555},
	publisher = {Optica Publishing Group},
	title = {Introduction to theory of high-harmonic generation in solids: tutorial},
	volume = {39},
	year = {2022}
}

@article{monkhorst1976special,
	author = {Monkhorst, Hendrik J and Pack, James D},
	doi = {10.1103/PhysRevB.13.5188},
	journal = {Physical review B},
	number = {12},
	pages = {5188},
	publisher = {APS},
	title = {Special points for Brillouin-zone integrations},
	volume = {13},
	year = {1976}
}

@article{wang2006ab,
	author = {Wang, Xinjie and Yates, Jonathan R and Souza, Ivo and Vanderbilt, David},
	doi = {10.1103/PhysRevB.74.195118},
	journal = {Physical Review B},
	number = {19},
	pages = {195118},
	publisher = {APS},
	title = {Ab initio calculation of the anomalous Hall conductivity by Wannier interpolation},
	volume = {74},
	year = {2006}
}

@article{ghimire2011observation,
	author = {Ghimire, Shambhu and DiChiara, Anthony D and Sistrunk, Emily and Agostini, Pierre and DiMauro, Louis F and Reis, David A},
	doi = {10.1038/nphys1847},
	journal = {Nature physics},
	number = {2},
	pages = {138--141},
	publisher = {Nature Publishing Group UK London},
	title = {Observation of high-order harmonic generation in a bulk crystal},
	volume = {7},
	year = {2011}
}

@article{kruchinin2018colloquium,
	author = {Kruchinin, Stanislav Yu and Krausz, Ferenc and Yakovlev, Vladislav S},
	doi = {10.1103/RevModPhys.90.021002},
	journal = {Reviews of Modern Physics},
	number = {2},
	pages = {021002},
	publisher = {APS},
	title = {Colloquium: Strong-field phenomena in periodic systems},
	volume = {90},
	year = {2018}
}

@Article{krausz2009attosecond,
  author    = {Krausz, Ferenc and Ivanov, Misha},
  title     = {Attosecond physics},
  journal   = {Reviews of Modern physics},
  year      = {2009},
  volume    = {81},
  number    = {1},
  pages     = {163--234},
  doi       = {10.1103/RevModPhys.81.163},
  publisher = {APS},
}

@article{ghimire2019high,
	author = {Ghimire, Shambhu and Reis, David A},
	doi = {10.1038/s41567-018-0315-5},
	journal = {Nature physics},
	number = {1},
	pages = {10--16},
	publisher = {Nature Publishing Group UK London},
	title = {High-harmonic generation from solids},
	volume = {15},
	year = {2019}
}

@article{cavaletto2025attoscience,
	author = {Cavaletto, Stefano M and Kowalczyk, Katarzyna M and Navarrete, Francisco O and Rivera-Dean, Javier},
	doi = {10.1038/s42254-024-00784-3},
	journal = {Nature Reviews Physics},
	number = {1},
	pages = {38--49},
	publisher = {Nature Publishing Group UK London},
	title = {The attoscience of strong-field-driven solids},
	volume = {7},
	year = {2025}
}

@article{heide2024ultrafast,
	author = {Heide, Christian and Kobayashi, Yuki and Haque, Sheikh Rubaiat Ul and Ghimire, Shambhu},
	doi = {10.1038/s41567-024-02640-8},
	journal = {Nature Physics},
	number = {10},
	pages = {1546--1557},
	publisher = {Nature Publishing Group UK London},
	title = {Ultrafast high-harmonic spectroscopy of solids},
	volume = {20},
	year = {2024}
}

@article{vampa2015all,
	author = {Vampa, Giulio and Hammond, TJ and Thir{\'e}, Nicolas and Schmidt, BE and L{\'e}gar{\'e}, Fran{\c{c}}ois and McDonald, CR and Brabec, Thomas and Klug, DD and Corkum, PB},
	doi = {10.1103/PhysRevLett.115.193603},
	journal = {Physical review letters},
	number = {19},
	pages = {193603},
	publisher = {APS},
	title = {All-optical reconstruction of crystal band structure},
	volume = {115},
	year = {2015}
}

@article{hohenleutner2015real,
	author = {Hohenleutner, Matthias and Langer, Fabian and Schubert, Olaf and Knorr, Matthias and Huttner, U and Koch, Stephan W and Kira, M and Huber, Rupert},
	doi = {10.1038/nature14652},
	journal = {Nature},
	number = {7562},
	pages = {572--575},
	publisher = {Nature Publishing Group UK London},
	title = {Real-time observation of interfering crystal electrons in high-harmonic generation},
	volume = {523},
	year = {2015}
}

@article{uzan2020attosecond,
	author = {Uzan, Ayelet Julie and Orenstein, Gal and Jim{\'e}nez-Gal{\'a}n, {\'A}lvaro and McDonald, Chris and Silva, Rui EF and Bruner, Barry D and Klimkin, Nikolai D and Blanchet, Valerie and Arusi-Parpar, Talya and Kr{\"u}ger, Michael and others},
	doi = {10.1038/s41566-019-0574-4},
	journal = {Nature Photonics},
	number = {3},
	pages = {183--187},
	publisher = {Nature Publishing Group UK London},
	title = {Attosecond spectral singularities in solid-state high-harmonic generation},
	volume = {14},
	year = {2020}
}

@article{silva2019topological,
	author = {Silva, REF and Jim{\'e}nez-Gal{\'a}n, {\'A} and Amorim, B and Smirnova, O and Ivanov, M},
	doi = {10.1038/s41566-019-0516-1},
	journal = {Nature Photonics},
	number = {12},
	pages = {849--854},
	publisher = {Nature Publishing Group UK London},
	title = {Topological strong-field physics on sub-laser-cycle timescale},
	volume = {13},
	year = {2019}
}

@article{silva2018high,
	author = {Silva, REF and Blinov, Igor V and Rubtsov, Alexey N and Smirnova, O and Ivanov, M},
	doi = {10.1038/s41566-019-0516-1},
	journal = {Nature Photonics},
	number = {5},
	pages = {266--270},
	publisher = {Nature Publishing Group UK London},
	title = {High-harmonic spectroscopy of ultrafast many-body dynamics in strongly correlated systems},
	volume = {12},
	year = {2018}
}

@article{murakami2018high,
	author = {Murakami, Yuta and Eckstein, Martin and Werner, Philipp},
	doi = {10.1103/PhysRevLett.121.057405},
	journal = {Physical review letters},
	number = {5},
	pages = {057405},
	publisher = {APS},
	title = {High-harmonic generation in Mott insulators},
	volume = {121},
	year = {2018}
}

@article{osika2017wannier,
	author = {Osika, Edyta N and Chac{\'o}n, Alexis and Ortmann, Lisa and Su{\'a}rez, Noslen and P{\'e}rez-Hern{\'a}ndez, Jose Antonio and Szafran, Bartłomiej and Ciappina, Marcelo F and Sols, Fernando and Landsman, Alexandra S and Lewenstein, Maciej},
	doi = {10.1103/PhysRevX.7.021017},
	journal = {Physical Review X},
	number = {2},
	pages = {021017},
	publisher = {APS},
	title = {Wannier-Bloch approach to localization in high-harmonics generation in solids},
	volume = {7},
	year = {2017}
}

@article{parks2020wannier,
	author = {Parks, Andrew M and Ernotte, Guilmot and Thorpe, Adam and McDonald, Chris R and Corkum, Paul B and Taucer, Marco and Brabec, Thomas},
	doi = {10.1364/OPTICA.402393},
	journal = {Optica},
	number = {12},
	pages = {1764--1772},
	publisher = {OSA},
	title = {Wannier quasi-classical approach to high harmonic generation in semiconductors},
	volume = {7},
	year = {2020}
}

@article{tancogne2017ellipticity,
	author = {Tancogne-Dejean, Nicolas and M{\"u}cke, Oliver D and K{\"a}rtner, Franz X and Rubio, Angel},
	doi = {10.1038/s41467-017-00764-5},
	journal = {Nature communications},
	number = {1},
	pages = {745},
	publisher = {Nature Publishing Group UK London},
	title = {Ellipticity dependence of high-harmonic generation in solids originating from coupled intraband and interband dynamics},
	volume = {8},
	year = {2017}
}

@article{Tancogne_Dejean_2018,
	author = {Tancogne-Dejean, Nicolas and Rubio, Angel},
	doi = {10.1126/sciadv.aao5207},
	issn = {2375-2548},
	journal = {Science Advances},
	number = {2},
	publisher = {American Association for the Advancement of Science (AAAS)},
	title = {Atomic-like high-harmonic generation from two-dimensional materials},
	volume = {4},
	year = {2018}
}

@article{lindberg1988effective,
	author = {Lindberg, M and Koch, Stephan W},
	doi = {10.1103/PhysRevB.38.3342},
	journal = {Physical Review B},
	number = {5},
	pages = {3342},
	publisher = {APS},
	title = {Effective Bloch equations for semiconductors},
	volume = {38},
	year = {1988}
}

@article{gu2022full,
	author = {Gu, J and Kolesik, M},
	doi = {10.1103/PhysRevA.106.063516},
	journal = {Physical Review A},
	number = {6},
	pages = {063516},
	publisher = {APS},
	title = {Full-Brillouin-zone calculation of high-order harmonic generation from solid-state media},
	volume = {106},
	year = {2022}
}

@article{Parks_2023,
	author = {Parks, A. M. and Moloney, J. V. and Brabec, T.},
	doi = {10.1103/physrevlett.131.236902},
	issn = {1079-7114},
	journal = {Physical Review Letters},
	number = {23},
	publisher = {American Physical Society (APS)},
	title = {Gauge Invariant Formulation of the Semiconductor Bloch Equations},
	volume = {131},
	year = {2023}
}

@article{yue2020structure,
	author = {Yue, Lun and Gaarde, Mette B},
	doi = {10.1103/PhysRevA.101.053411},
	journal = {Physical Review A},
	number = {5},
	pages = {053411},
	publisher = {APS},
	title = {Structure gauges and laser gauges for the semiconductor Bloch equations in high-order harmonic generation in solids},
	volume = {101},
	year = {2020}
}

@article{Yakovlev_2017,
	author = {Yakovlev, Vladislav S. and Wismer, Michael S.},
	doi = {10.1016/j.cpc.2017.04.010},
	issn = {0010-4655},
	journal = {Computer Physics Communications},
	pages = {82--88},
	publisher = {Elsevier BV},
	title = {Adiabatic corrections for velocity-gauge simulations of electron dynamics in periodic potentials},
	volume = {217},
	year = {2017}
}

@article{Bertolino_2022,
	author = {Bertolino, Mattias and Carlstr{\"o}m, Stefanos and Peschel, Jasper and Zapata, Felipe and Lindroth, Eva and Dahlstr{\"o}m, Jan Marcus},
	doi = {10.1103/physreva.106.043108},
	issn = {2469-9934},
	journal = {Physical Review A},
	number = {4},
	publisher = {American Physical Society (APS)},
	title = {Thomas--Reiche--Kuhn correction for truncated configuration-interaction spaces: Case of laser-assisted dynamical interference},
	volume = {106},
	year = {2022}
}

@article{Kobe_1979,
	author = {Kobe, Donald H.},
	doi = {10.1103/physreva.19.205},
	issn = {0556-2791},
	journal = {Physical Review A},
	number = {1},
	pages = {205--214},
	publisher = {American Physical Society (APS)},
	title = {Gauge-invariant resolution of the controversy over length versus velocity forms of the interaction with electric dipole radiation},
	volume = {19},
	year = {1979}
}

@article{Parks_2024,
	author = {Parks, A. M. and Moloney, J. V. and Brabec, T.},
	doi = {10.1364/josab.520221},
	issn = {1520-8540},
	journal = {Journal of the Optical Society of America B},
	number = {6},
	pages = {B47},
	publisher = {Optica Publishing Group},
	title = {Bloch gauge symmetry of the semiconductor Bloch equations [Invited]},
	volume = {41},
	year = {2024}
}

@article{Jiang_2018,
	author = {Jiang, Shicheng and Chen, Jigen and Wei, Hui and Yu, Chao and Lu, Ruifeng and Lin, C. D.},
	doi = {10.1103/physrevlett.120.253201},
	issn = {1079-7114},
	journal = {Physical Review Letters},
	number = {25},
	publisher = {American Physical Society (APS)},
	title = {Role of the Transition Dipole Amplitude and Phase on the Generation of Odd and Even High-Order Harmonics in Crystals},
	volume = {120},
	year = {2018}
}

@article{corkum1993plasma,
	author = {Corkum, Paul B},
	doi = {10.1103/physrevlett.71.1994},
	journal = {Physical review letters},
	number = {13},
	pages = {1994},
	publisher = {APS},
	title = {Plasma perspective on strong field multiphoton ionization},
	volume = {71},
	year = {1993}
}

@article{Krause_1992,
	author = {Krause, Jeffrey L. and Schafer, Kenneth J. and Kulander, Kenneth C.},
	doi = {10.1103/physrevlett.68.3535},
	issn = {0031-9007},
	journal = {Physical Review Letters},
	number = {24},
	pages = {3535--3538},
	publisher = {American Physical Society (APS)},
	title = {High-order harmonic generation from atoms and ions in the high intensity regime},
	volume = {68},
	year = {1992}
}

@article{Lewenstein_1994,
	author = {Lewenstein, M. and Balcou, Ph. and Ivanov, M. Yu. and L'Huillier, Anne and Corkum, P. B.},
	doi = {10.1103/physreva.49.2117},
	issn = {1094-1622},
	journal = {Physical Review A},
	number = {3},
	pages = {2117--2132},
	publisher = {American Physical Society (APS)},
	title = {Theory of high-harmonic generation by low-frequency laser fields},
	volume = {49},
	year = {1994}
}

@inbook{doi:https://doi.org/10.1002/9783527677689.ch7,
	author = {Smirnova, Olga and Ivanov, Misha},
	booktitle = {Attosecond and XUV Physics},
	chapter = {7},
	doi = {10.1002/9783527677689.ch7},
	eprint = {https://onlinelibrary.wiley.com/doi/pdf/10.1002/9783527677689.ch7},
	isbn = {9783527677689},
	pages = {201--256},
	publisher = {John Wiley \& Sons, Ltd},
	title = {Multielectron High Harmonic Generation: Simple Man on a Complex Plane},
	url = {https://onlinelibrary.wiley.com/doi/abs/10.1002/9783527677689.ch7},
	year = {2014}
}

@article{amini2019symphony,
	author = {Amini, Kasra and Biegert, Jens and Calegari, Francesca and Chac{\'o}n, Alexis and Ciappina, Marcelo F and Dauphin, Alexandre and Efimov, Dmitry K and de Morisson Faria, Carla Figueira and Giergiel, Krzysztof and Gniewek, Piotr and others},
	doi = {10.1088/1361-6633/ab2bb1},
	journal = {Reports on Progress in Physics},
	number = {11},
	pages = {116001},
	publisher = {IOP Publishing},
	title = {Symphony on strong field approximation},
	volume = {82},
	year = {2019}
}

@article{Lara_Astiaso_2016,
	author = {Lara-Astiaso, M. and Silva, R. E. F. and Gubaydullin, A. and Rivi{\`e}re, P. and Meier, C. and Mart{\'i}n, F.},
	doi = {10.1103/physrevlett.117.093003},
	issn = {1079-7114},
	journal = {Physical Review Letters},
	number = {9},
	publisher = {American Physical Society (APS)},
	title = {Enhancing High-Order Harmonic Generation in Light Molecules by Using Chirped Pulses},
	volume = {117},
	year = {2016}
}

@article{Van_Tuan_2018,
	author = {{Van Tuan}, Dinh and Yang, Min and Dery, Hanan},
	doi = {10.1103/physrevb.98.125308},
	issn = {2469-9969},
	journal = {Physical Review B},
	number = {12},
	publisher = {American Physical Society (APS)},
	title = {Coulomb interaction in monolayer transition-metal dichalcogenides},
	volume = {98},
	year = {2018}
}

@article{rytova2018screened,
	author = {Rytova, Natalia S},
	doi = {10.48550/arXiv.1806.00976},
	journal = {arXiv preprint arXiv:1806.00976},
	title = {Screened potential of a point charge in a thin film},
	year = {2018}
}

@article{keldysh1979coulomb,
	author = {Keldysh, LV},
	journal = {Zh. Eksp. Teor. Fiz., Pis' ma},
	number = {11},
	pages = {716--719},
	title = {Coulomb interaction in thin films of semiconductors and semimetals},
	volume = {29},
	year = {1979}
}

@article{Wu_2015,
	author = {Wu, Fengcheng and Qu, Fanyao and MacDonald, A. H.},
	doi = {10.1103/physrevb.91.075310},
	issn = {1550-235X},
	journal = {Physical Review B},
	number = {7},
	publisher = {American Physical Society (APS)},
	title = {Exciton band structure of monolayer MoS2},
	volume = {91},
	year = {2015}
}

@article{uria2024efficient,
	author = {Ur{\'i}a-{\'A}lvarez, Alejandro Jos{\'e} and Esteve-Paredes, Juan Jos{\'e} and Garc{\'i}a-Bl{\'a}zquez, Manuel Antonio and Palacios, Juan Jos{\'e}},
	doi = {10.1016/j.cpc.2023.109001},
	journal = {Computer Physics Communications},
	pages = {109001},
	publisher = {Elsevier},
	title = {Efficient computation of optical excitations in two-dimensional materials with the Xatu code},
	volume = {295},
	year = {2024}
}

@article{prodan2005nearsightedness,
	author = {Prodan, Emil and Kohn, Walter},
	doi = {10.1073/pnas.0505436102},
	journal = {Proceedings of the National Academy of Sciences},
	number = {33},
	pages = {11635--11638},
	publisher = {National Academy of Sciences},
	title = {Nearsightedness of electronic matter},
	volume = {102},
	year = {2005}
}

@article{Goedecker_1999,
	author = {Goedecker, Stefan},
	doi = {10.1103/revmodphys.71.1085},
	issn = {1539-0756},
	journal = {Reviews of Modern Physics},
	number = {4},
	pages = {1085--1123},
	publisher = {American Physical Society (APS)},
	title = {Linear scaling electronic structure methods},
	volume = {71},
	year = {1999}
}

@article{Kohn_1996,
	author = {Kohn, W.},
	doi = {10.1103/physrevlett.76.3168},
	issn = {1079-7114},
	journal = {Physical Review Letters},
	number = {17},
	pages = {3168--3171},
	publisher = {American Physical Society (APS)},
	title = {Density Functional and Density Matrix Method Scaling Linearly with the Number of Atoms},
	volume = {76},
	year = {1996}
}

@article{Goedecker_1998,
	author = {Goedecker, S.},
	doi = {10.1103/physrevb.58.3501},
	issn = {1095-3795},
	journal = {Physical Review B},
	number = {7},
	pages = {3501--3502},
	publisher = {American Physical Society (APS)},
	title = {Decay properties of the finite-temperature density matrix in metals},
	volume = {58},
	year = {1998}
}

@article{prentice2020onetep,
	author = {Prentice, Joseph CA and Aarons, Jolyon and Womack, James C and Allen, Alice EA and Andrinopoulos, Lampros and Anton, Lucian and Bell, Robert A and Bhandari, Arihant and Bramley, Gabriel A and Charlton, Robert J and others},
	doi = {10.1063/5.0004445},
	journal = {The Journal of chemical physics},
	number = {17},
	publisher = {AIP Publishing},
	title = {The ONETEP linear-scaling density functional theory program},
	volume = {152},
	year = {2020}
}

@article{castro2009electronic,
	author = {{Castro Neto}, Antonio H and Guinea, Francisco and Peres, Nuno MR and Novoselov, Kostya S and Geim, Andre K},
	doi = {10.1103/RevModPhys.81.109},
	journal = {Reviews of modern physics},
	number = {1},
	pages = {109--162},
	publisher = {APS},
	title = {The electronic properties of graphene},
	volume = {81},
	year = {2009}
}

@article{lobo2024exponential,
	author = {Lobo, Francisco and Prada, Elsa and San-Jose, Pablo},
	doi = {10.1103/PhysRevB.111.195415},
	journal = {Physical Review B},
	number = {19},
	pages = {195415},
	publisher = {APS},
	title = {Exponential suppression of the topological gap in self-consistent intrinsic Majorana nanowires},
	volume = {111},
	year = {2025}
}

@article{ridolfi2020expeditious,
	author = {Ridolfi, Emilia and Trevisanutto, Paolo E and Pereira, Vitor M},
	doi = {10.1103/PhysRevB.102.245110},
	journal = {Physical Review B},
	number = {24},
	pages = {245110},
	publisher = {APS},
	title = {Expeditious computation of nonlinear optical properties of arbitrary order with native electronic interactions in the time domain},
	volume = {102},
	year = {2020}
}

@article{heyd2003hybrid,
	author = {Heyd, Jochen and Scuseria, Gustavo E and Ernzerhof, Matthias},
	doi = {10.1063/1.1564060},
	journal = {The Journal of chemical physics},
	number = {18},
	pages = {8207--8215},
	publisher = {American Institute of Physics},
	title = {Hybrid functionals based on a screened Coulomb potential},
	volume = {118},
	year = {2003}
}

@Article{boroumand2025strong,
  author    = {Boroumand, Neda and Thorpe, Adam and Bart, Graeme and Parks, Andrew M and Toutounji, Mohamad and Vampa, Giulio and Brabec, Thomas and Wang, Lu},
  title     = {Strong field physics in open quantum systems},
  journal   = {Reports on Progress in Physics},
  year      = {2025},
  volume    = {88},
  number    = {7},
  pages     = {070501},
  doi       = {10.1088/1361-6633/adeebb},
  publisher = {IOP Publishing},
}

@Article{tumakov2026real,
  author  = {Tumakov, Dmitry and Popova-Gorelova, Daria},
  title   = {Real-time exciton dynamics in two-dimensional materials under ultrashort laser pulses},
  journal = {arXiv preprint arXiv:2603.06446},
  year    = {2026},
  doi     = {10.48550/arXiv.2603.06446},
}

@PhdThesis{molinero2025phd,
  author = {Eduardo Bernal Molinero},
  title  = {Strong field physics in periodic crystals},
  school = {Universidad Autónoma de Madrid},
  year   = {2025},
  url    = {http://hdl.handle.net/10486/719985},
}
\end{document}